\documentclass[aps,reprint,superscriptaddress,longbibliography]{revtex4-2}
\usepackage{qaoa_manuscript}
\usepackage[normalem]{ulem}
\hypersetup{pdfauthor={Myeongsu Kim and Sabre Kais},pdftitle={Physical Response and Predictive Information in Quantum Approximate Optimization}}
\begin{document}
\title{Physical Response and Predictive Information in Quantum Approximate Optimization}
\author{Myeongsu Kim}
\affiliation{Department of Computer Science, Purdue University, West Lafayette, IN, USA}
\author{Sabre Kais}
\email[Corresponding author: ]{skais@ncsu.edu}
\affiliation{Department of Electrical and Computer Engineering, North Carolina State University, Raleigh, NC 27695, USA}
\affiliation{Department of Chemistry, North Carolina State University, Raleigh, NC 27695, USA}
\begin{abstract}
Which features of a quantum state make a compact observable useful for predicting its later response? We address this question through the quadratic response of good-solution probability to a phase intervention in the quantum approximate optimization algorithm. Exact decompositions of the response and prediction loss relate the benefit of three observable means to errors left by a strong baseline. Across independent evaluations at 10--16 qubits, the prediction changes produced by these observables track errors in direct or complementary response components; opposing alignments can limit the net gain. In Max3SAT with analytically initialized preparation and continuations centred on calibrated schedules, the summaries reduce response-prediction mean squared error by 19.20--34.56\%. In Max3XOR, prediction changes from adding low-order summaries align with higher-weight response residuals, even though the summaries' own operator subspace contributes little response. Paired preparations under a shared response operator give different alignment profiles, while matched-state continuation controls show how subsequent dynamics reshape usefulness. Predictive gains also persist in high-reward preparations and with frozen predictors on fresh problems at unseen angle centres within the trained range. These results provide a physical basis for designing compact state representations around the target response and the information already available.
\end{abstract}
\maketitle

\section{Introduction}

Predicting how a quantum state responds to a later circuit requires a useful description of that state. Compact observables provide such descriptions without reconstructing the full wavefunction, but their value for prediction depends on more than the response they carry directly. The quantum approximate optimization algorithm (QAOA) provides a concrete setting for studying this relationship: alternating cost and mixing operations prepare candidate solutions, and the state created along the way determines the effect of subsequent operations~\cite{farhi2014,zhou2020,capecci2025}.

Direct response contribution measures how a specified operator component enters the physical response. Additional predictive value measures how much its observables reduce prediction error after the available baseline information has been supplied. A large direct component may add little predictive value if the baseline already resolves it or if opposing residual alignments suppress the net improvement. Conversely, observables with a small direct contribution may track response components outside their own operator subspace through correlations in the prepared-state ensemble. Connecting these two quantities requires a physical decomposition tied to the prediction error that remains after known information has been accounted for.

Observable-based quantum prediction and control motivate this question from complementary directions. Intermediate QAOA observables inform reinforcement learning~\cite{garciasaez2019,wauters2020}, while measured commutators guide feedback-based quantum optimization~\cite{magann2022}. Quantum learning links local measurements to later dynamics and short-range correlations to global properties~\cite{huang2023,wu2024}; intermediate structure also predicts rare outcomes in local random circuits~\cite{kim2026predictive}. Within QAOA, coherence-based descriptions of initial states~\cite{caliz2025}, prediction from graph and circuit features~\cite{coelho2026}, and commutator expansions of observables and probabilities~\cite{hadfield2023} provide related perspectives.

Here we connect intermediate observables' direct response contributions to their additional predictive value. Exact response and prediction-loss decompositions identify direct-component alignment, opposing residual alignments and complementary information in independently evaluated Max3SAT and Max3XOR ensembles at $n=10,12,14,16$. Resolving the XOR complement by Pauli weight locates the higher-weight residuals associated with low-order summaries and compares paired preparations under a shared response operator. High-reward preparation and fixed-state continuation controls distinguish the roles of the prepared ensemble and subsequent dynamics, while frozen predictors test usefulness on fresh problems at unseen angle centres.

\section{Results}

\subsection{Phase sensitivity and compact state observables}

We use the same signed three-variable clause instances, with two clauses per variable, for Max3SAT (OR) and Max3XOR (XOR). This common clause structure provides a setting for examining compact state observables under different cost functions. A reward $R(z)$ counts the satisfied clauses of a bitstring $z$. We define good solutions through the attained reward range,
\begin{equation}
 R(z)\geq R_{\max}-0.05(R_{\max}-R_{\min}),
 \label{eq:good_threshold}
\end{equation}
with the numerical tolerance specified in Methods. Thus, the good set is determined by reward rather than by the number of bitstrings it contains.

A depth-$p\in\{4,8\}$ QAOA circuit is divided into a preparation prefix and a continuation of $p/2$ layers each. A and C denote analytic and separately calibrated schedule centres; the two-letter labels specify preparation first and continuation second. Preparation uses instance-optimized angles initialized near the indicated centre, whereas continuations are sampled around fixed centres (Methods~\ref{meth:circuits}). Thus AA and AC share a prepared state within each matched instance and isolate a change in subsequent dynamics; CC uses C for both roles.

We insert $e^{-idG}$ between the prefix and a continuation $V_\theta$. The diagonal generator $G$ is a normalized linear combination of either single-qubit $Z$ operators or two-qubit $ZZ$ operators. With $\Pi_{\mathrm{good}}$ projecting onto Eq.~\eqref{eq:good_threshold}, the output probability is
\begin{align}
 P_\theta(d)
 &=\langle\psi|e^{idG}V_\theta^\dagger
       \Pi_{\mathrm{good}}V_\theta e^{-idG}|\psi\rangle \\
 &=P_\theta(0)+c_{1,\theta}d+c_{2,\theta}d^2+O(d^3).
 \label{eq:response_series}
\end{align}
Our target $c_2$ is the average of $c_{2,\theta}=P_\theta''(0)/2$ over 64 sampled continuations. It governs the leading change in good-solution probability under a small symmetric phase kick. Writing $\overline P$ for the same continuation average gives
\begin{equation}
 \frac{\overline P(\epsilon)+\overline P(-\epsilon)}{2}-\overline P(0)
 =\epsilon^2c_2+O(\epsilon^4).
 \label{eq:symmetric_response}
\end{equation}
Exact statevector calculations supply the coefficients without finite-difference differentiation. The continuation samples describe variation in the future circuit.

The baseline input $B$ includes cost mean and variance, computational-basis entropy, a reward histogram, all single-site $Z$ and pairwise $ZZ$ expectations, responses to mixer probes, and descriptors of the instance, preparation, generator and continuation. The added summaries are
\begin{equation}
\begin{aligned}
s_X&=\frac{1}{n}\sum_i\langle X_i\rangle,\\
 s_{XX}&=\frac{1}{|E|}\sum_{(i,j)\in E}\langle X_iX_j\rangle,\\
 s_{YY}&=\frac{1}{|E|}\sum_{(i,j)\in E}\langle Y_iY_j\rangle,
\end{aligned}
 \label{eq:three_summaries}
\end{equation}
where $E$ contains the distinct pairs occurring in the clauses. The summary $s_X$ is the normalized expectation of the QAOA mixer, while $s_{XX}$ and $s_{YY}$ capture distinct off-diagonal correlations on clause-supported pairs. Averaging over sites and pairs yields three scalar features at every system size; we denote their joint addition by M3. These summaries can change under diagonal phase transformations that preserve the computational-basis distribution. Their predictive value therefore tests compact off-diagonal state information beyond the cost, probability statistics and mixer-response features already supplied by the baseline.

\begin{figure*}[tbp]
 \centering
 \manuscriptfigure{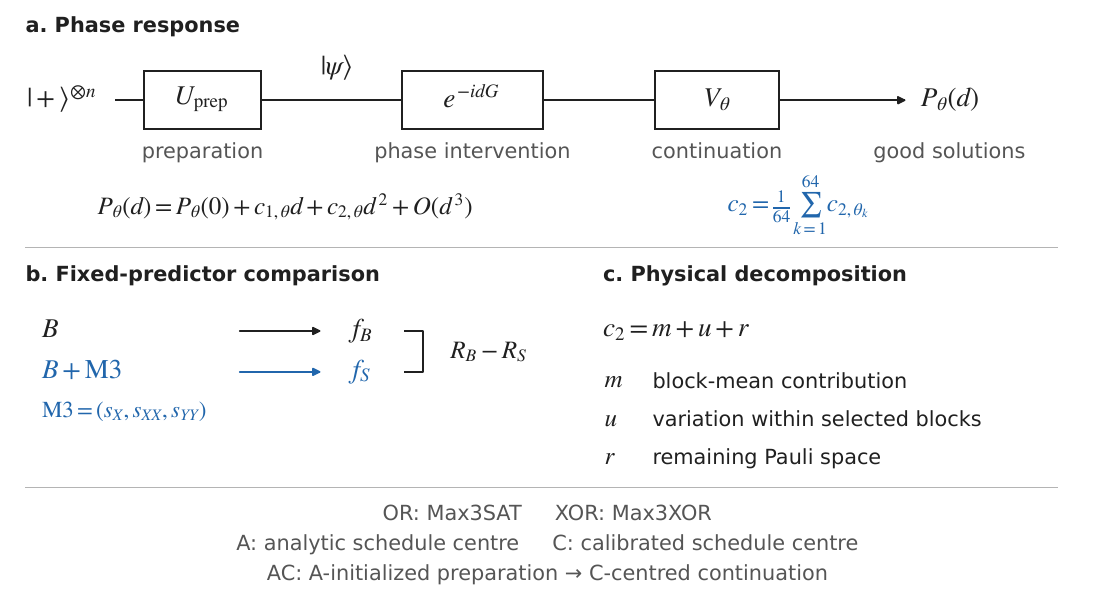}
 \caption{\textbf{Diagnosing the predictive information in intermediate observables.}
 \textbf{a,} A diagonal phase intervention follows QAOA state preparation and precedes a sampled continuation. The target $c_2$ is the quadratic coefficient of good-solution probability, averaged over 64 continuations. Preparation includes the local-$Z$ transformation where specified.
 \textbf{b,} Fixed predictors using baseline information $B$ or $B$ augmented with $\mathrm{M3}=(s_X,s_{XX},s_{YY})$ are compared on the same evaluation instances; $R_B$ and $R_S$ denote their mean squared errors.
 \textbf{c,} The exact decomposition separates the block-mean contribution $m$, the contribution $u$ from nonuniform coefficients within the selected blocks, and the remaining Pauli-space contribution $r$.
 The four preparation--continuation conditions are OR AA, OR AC, OR AC with a local-$Z$ transformation, and XOR CC. At each of $n=10,12,14,16$, 256 instances are used for training and 512 independent instances for evaluation. Depths $p=4,8$ and $Z/ZZ$ generators are repeated observations within an instance. Study designs and statistical comparisons are specified in Supplementary Table~\ref{supp:studyguide} and Methods~\ref{meth:statistics}.}
 \label{fig:design}
\end{figure*}

The core exact-component comparison covers OR AA, OR AC, a local-$Z$ transformation of OR AC, and XOR CC at $n=10,12,14,16$, using predictors fixed before independent evaluation (Fig.~\ref{fig:design}). The matched OR conditions probe how continuation and phase transformations change the summaries' usefulness; XOR tests the role of information beyond the summaries' own response contribution. The study designs and their separate inference families are summarized in Supplementary Table~\ref{supp:studyguide} and Methods~\ref{meth:statistics}.

\subsection{Compact observables improve predictions on new instances}

Let $f_B$ and $f_S$ denote the scalar predictions with the baseline and augmented inputs, respectively. We report the relative reduction in mean squared error,
\begin{equation}
 g=\frac{R_B-R_S}{R_B},\qquad
 R_v=\mathbb{E}[(c_2-f_v)^2].
 \label{eq:gain}
\end{equation}
The expectation weights independent instances equally after averaging their repeated depth--generator conditions. Throughout, predictive value and additional predictive information denote held-out error reduction by the fitted predictors relative to the specified baseline.

\begin{figure*}[tbp]
 \centering
 \manuscriptfigure[0.64\textheight]{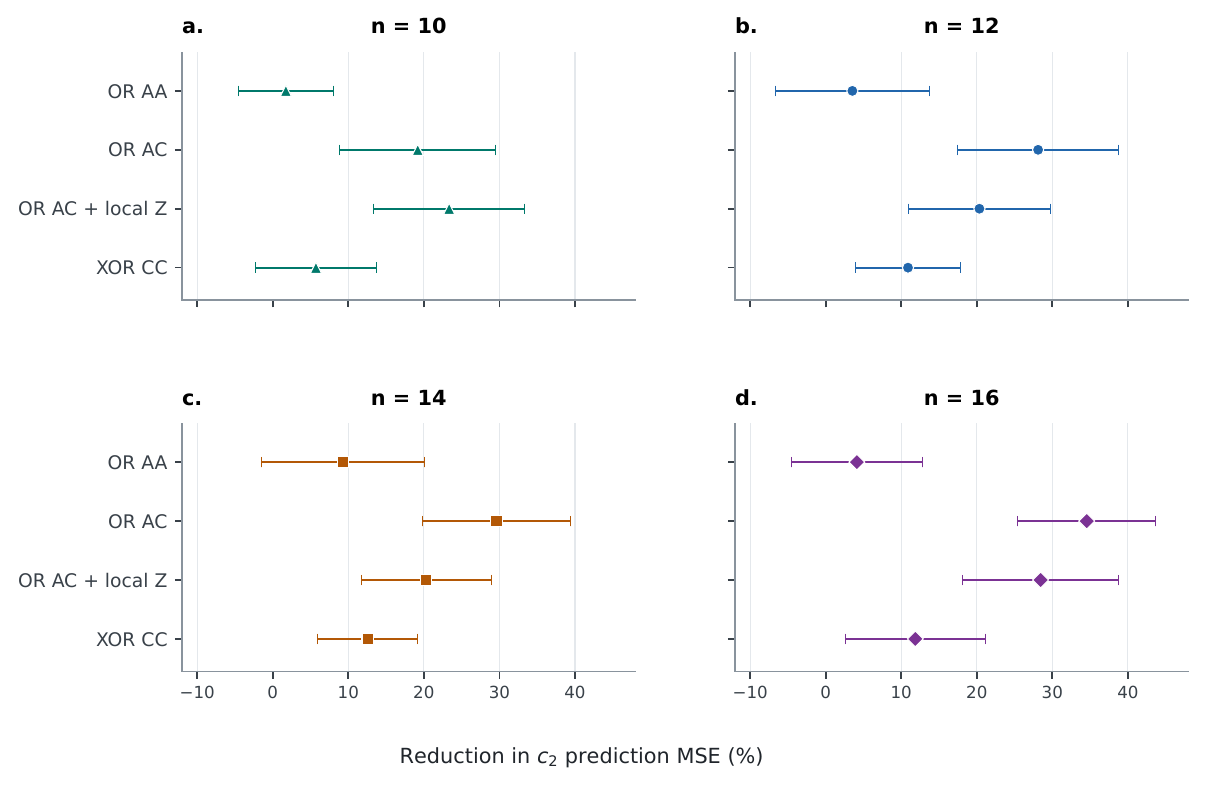}
 \caption{\textbf{Additional prediction accuracy on independent problems.}
 Relative reduction $100g$ in the MSE of the complete response $c_2$ after adding M3 to the Extra Trees baseline, at (a) $n=10$, (b) $n=12$, (c) $n=14$ and (d) $n=16$. Each estimate uses 512 evaluation instances, with depth and generator repeats grouped within the instance. Intervals are 95\% simultaneous within separate secondary families for $n=10$, joint $n=12,14$, and $n=16$ (Methods~\ref{meth:statistics}). OR AC has positive bounds at all four sizes; the $n=10$ XOR interval in this exact-component study includes zero. Its separate sample-robustness evaluation appears in Supplementary Table~\ref{tab:n10-robustness}.}
 \label{fig:gain}
\end{figure*}

Across $n=10,12,14,16$, the prespecified Extra Trees/M3 model reduces OR AC MSE by 19.20\%, 28.12\%, 29.63\% and 34.56\%, respectively, with positive simultaneous bounds at every size (Fig.~\ref{fig:gain}). XOR CC gives gains of 5.72\%, 10.89\%, 12.56\% and 11.87\%; the bounds are positive at $n=12,14,16$, while the $n=10$ interval includes zero. OR AA has smaller point estimates of 1.74--9.32\%, whose intervals include zero.

A separate $n=10$ evaluation on 1,024 fresh instances gives positive XOR gains for the same frozen model and after pooling losses across models trained on five independent cohorts (Supplementary Table~\ref{tab:n10-robustness}).

The OR AC benefit also persists at $n=14,16$ when the calibration budget is matched to that used at $n=10,12$. Across three independently calibrated C schedule sets at each size, gains are 26.11--29.87\% at $n=14$ and 33.12--33.49\% at $n=16$ on fresh training and evaluation cohorts, with positive 95\% simultaneous bounds for all six comparisons (Supplementary Section~\ref{supp:calibrationrobustness}). Ridge/M3 gives positive gain estimates in OR AC and XOR CC at every size, though smaller than Extra Trees. Adding Z2---the mean single-site $Z$ and clause-supported $ZZ$ expectations---to the baseline also leaves positive simultaneous bounds for M3 in OR AC at every size and XOR CC at $n=12,14,16$ (Supplementary Section~\ref{supp:diagonalcontrol}).

Comparisons with reduced observable sets assess how much of M3's predictive benefit is retained by individual or paired summaries (Methods~\ref{meth:learning}); additional cohort designs are summarized in Supplementary Section~\ref{supp:cohorts}. We next ask why the same observable set provides different benefits under different preparation and continuation conditions.

\subsection{Connecting the response operator to actual prediction gains}

To explain the different gains, we first identify which parts of the physical response the summaries carry. For each continuation define
\begin{equation}
 A_\theta=V_\theta^\dagger\Pi_{\mathrm{good}}V_\theta,
 \qquad K_\theta=-\tfrac12[G,[G,A_\theta]].
 \label{eq:response_operator}
\end{equation}
Then $c_{2,\theta}=\langle\psi|K_\theta|\psi\rangle$: $K_\theta$ specifies how the continuation converts the intermediate state into a quadratic response. This is the standard commutator representation of the quadratic response~\cite{hadfield2023}. We use its Pauli expansion to relate the target to the chosen summaries. Since $G$ is diagonal, $K_\theta$ has zero diagonal entries, so all purely Z-type Pauli coefficients vanish. The baseline's Z/ZZ observables can therefore predict this response through ensemble correlations despite having zero direct coefficients. Its mixer-probe features also supply phase-sensitive information (Methods~\ref{meth:components}).

Let $\mathcal{P}$ contain the $X_i$, $X_iX_j$ and $Y_iY_j$ operators entering Eq.~\eqref{eq:three_summaries}. Write $\alpha_P=2^{-n}\operatorname{Tr}(K_\theta P)$ and $e_P=\langle P\rangle$. Within each of the three blocks, separate the coefficient into its mean and its deviation from that mean. This gives the exact decomposition
\begin{equation}
 c_2=m+u+r.
 \label{eq:component_sum}
\end{equation}
Here $m$ is the response contribution obtained by replacing the coefficients in each selected block by their mean. The term $u$ restores the contribution from deviations around those means, and $r$ is the contribution outside $\mathcal{P}$. We refer to $m$ as the block-mean direct component; $m+u$ is the full contribution from the selected Pauli subspace. The same definitions apply after averaging over continuations. Exact full-rank operator coefficients are used at all four sizes. These coefficients and the component labels are diagnostic quantities, not inputs to the response predictors.

To explain the prediction gain, we compare the change from adding M3 with what the baseline misses in each response component. Write $\delta=f_S-f_B$, and fit a separate three-output component predictor $(g_m,g_u,g_r)$ using only $B$. With $d_B=f_B-g_m-g_u-g_r$, the actual scalar loss difference satisfies
\par
\Needspace{14\baselineskip}
\begin{samepage}
\begin{equation}
\begin{aligned}
R_B-R_S&=2\mathbb{E}[(m-g_m)\delta]
 +2\mathbb{E}[(u-g_u)\delta]\\
 &+2\mathbb{E}[(r-g_r)\delta]\\
 &-2\mathbb{E}[d_B\delta]-\mathbb{E}[\delta^2].
\end{aligned}
 \label{eq:loss_identity}
\end{equation}
We call the first three terms $T_m,T_u,T_r$. They measure signed alignment between the change in scalar prediction and the component residuals left by the baseline. The fourth term corrects the difference between the separately fitted scalar and component predictors, and the fifth penalizes the squared prediction change.\par
\end{samepage}

Retaining all five terms recovers the measured gain exactly, linking prediction improvement to unresolved component residuals rather than to component magnitudes alone. The identity holds for any such fitted predictors; the signs and magnitudes of its terms are empirical properties of the predictors and state ensemble. Their individual values depend on the fitted component predictors and are not causal fractions of the gain.

\subsection{Residual alignment and physical cancellation distinguish the OR conditions}

\begin{figure*}[tbp]
 \centering
 \manuscriptfigure[0.73\textheight]{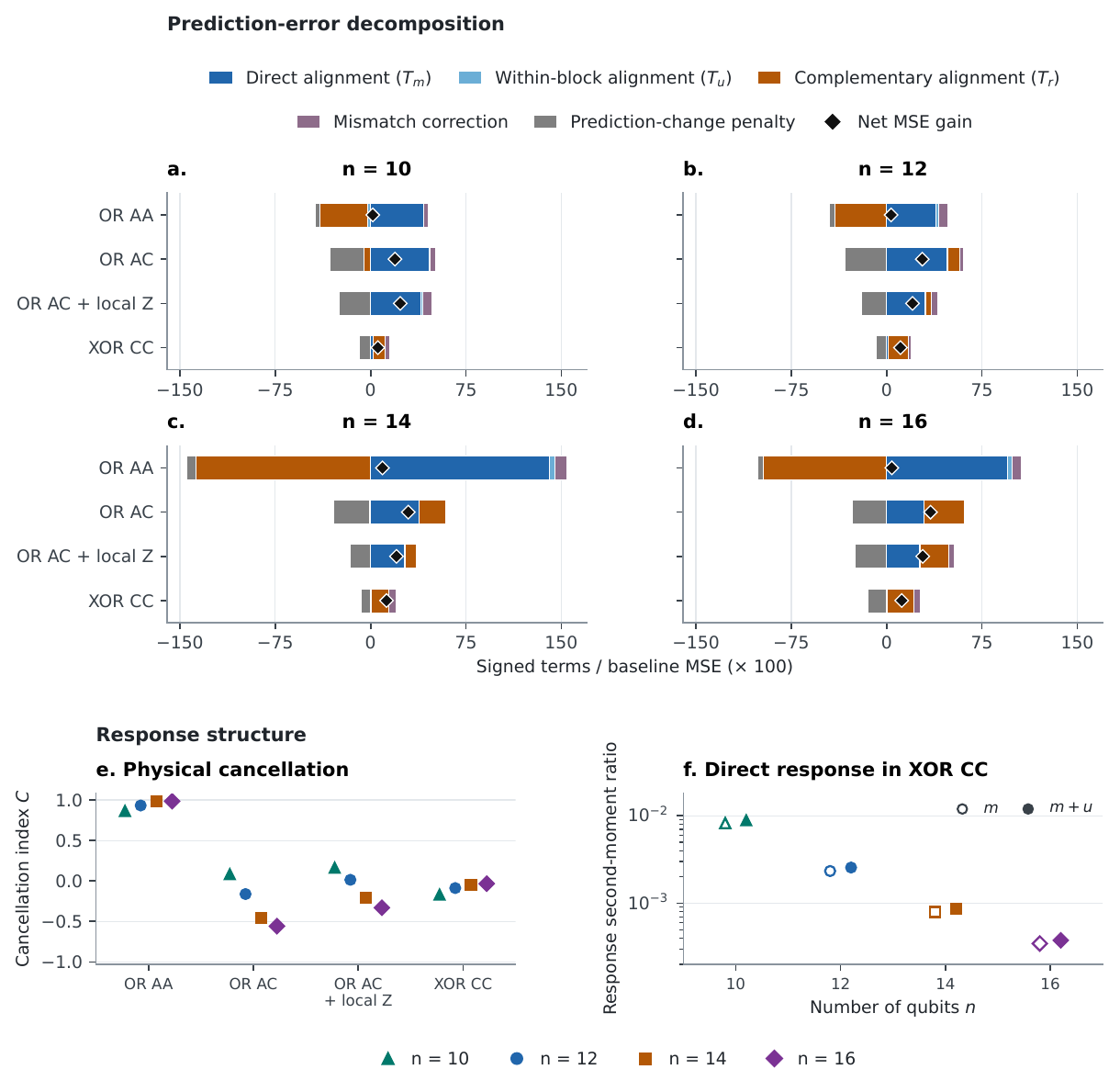}
 \caption{\textbf{Residual alignment, physical cancellation and complementary information.}
 (a--d) The five terms in Eq.~\eqref{eq:loss_identity} for $n=10,12,14,16$, respectively, divided by the condition-specific baseline MSE and multiplied by 100. Positive and negative terms are stacked separately on a common horizontal scale; black diamonds give their sum, the actual MSE reduction. Individual signed terms can exceed 100 when their opposing contributions offset in the net gain. (e) The physical cancellation index $C$ in Eq.~\eqref{eq:cancellation}. (f) Second-moment ratios in the exact-component XOR CC cohorts for the block-mean response $m$ (open symbols) and full selected-space response $m+u$ (filled symbols), each divided by $\mathbb{E}[c_2^2]$, on a logarithmic axis. Panels (e,f) show point estimates from the same evaluation cohorts as panels (a--d). OR AA combines opposing terms, OR AC retains positive direct alignment with an ensemble-dependent complementary term, and XOR CC has a small direct component alongside a positive complementary term. Each size retains its fixed calibration. OR--XOR comparisons change both predicate and preparation.}
 \label{fig:accounts}
\end{figure*}

The matched OR conditions show how direct-component information can yield different net benefits. In OR AC, $T_m$ has positive simultaneous bounds at all four sizes (Fig.~\ref{fig:accounts}; Supplementary Fig.~\ref{supp:confirmation}). The normalized direct-alignment term $100T_m/R_B$ ranges from 28.95 to 47.50, showing that the added observables move predictions in a direction associated with the direct-component residual. Complementary alignment varies across the ensembles: $100T_r/R_B$ is $-5.17$ at $n=10$ and 31.30 at $n=16$, where it is comparable to the positive direct term. OR AC therefore remains informative across different balances of direct and complementary response information.

OR AA combines positive direct alignment with an opposing complementary alignment at all four sizes, with simultaneous intervals supporting both signs. Both terms have magnitudes much larger than the net gain. Their opposition, together with the other terms in Eq.~\eqref{eq:loss_identity}, leaves the smaller scalar improvements in Fig.~\ref{fig:gain}. The full signed account in Fig.~\ref{fig:accounts} makes this difference between component alignment and realized predictive benefit explicit.

We separately characterize physical cancellation using only the exact response components, independently of the predictors:
\begin{equation}
 C=\frac{-2\mathbb{E}[m(u+r)]}
         {\mathbb{E}[m^2]+\mathbb{E}[(u+r)^2]}.
 \label{eq:cancellation}
\end{equation}
Here the italic $C$ denotes the cancellation index, while C in the recipe labels denotes the calibrated centre. For a positive denominator, this bounded index is close to one when the two contributions oppose one another strongly in their uncentred second moments. It equals $1-\mathbb{E}[c_2^2]/\{\mathbb{E}[m^2]+\mathbb{E}[(u+r)^2]\}$, and quantifies cancellation between the direct component and the combined remaining components. OR AA has consistently strong cancellation, $C=0.864$--0.986, whereas OR AC ranges from $-0.559$ to 0.083. All four AA--AC differences have positive simultaneous bounds. The recurring relationship is stronger cancellation in AA; the smaller AC index changes sign across the fixed calibrated ensembles. This recipe dependence appears both in the physical components and in their signed association with prediction changes.

After removing the overall means, AA retains strong cancellation and AC has a negative index at all four sizes (Supplementary Table~\ref{supp:centeringtable}). Removing the mean separately within each depth--generator condition gives fluctuation-cancellation indices of 0.916--0.983 for AA and $-0.308$ to $-0.061$ for AC across $n=10,12,14,16$. The pooled and within-condition analyses therefore support the same AA--AC distinction. Centering changes the magnitudes, especially in AC, so the original second-moment index and the fluctuation index describe different aspects of the ensemble.

A local-$Z$ phase control probes this direct association while preserving the OR AC state's computational-basis probabilities, diagonal cost and $Z/ZZ$ expectations. The prespecified identity-minus-rotation contrasts in $T_m$ have positive simultaneous bounds at $n=10,12,16$ (Supplementary Fig.~\ref{supp:confirmation}), confirming attenuation of direct-component alignment. The common identity baseline MSE sets the normalization. Mixer-related baseline features are recomputed and can change, so the control holds the diagonal probability information fixed. M3 remains useful after the rotation, with MSE reductions of 20.33--28.44\% across the four sizes.

\subsection{Low-order observables carry information about higher-weight response}

XOR CC addresses the complementary question: can these summaries improve prediction when their selected operator subspace carries very little response? In the exact-component cohorts, the full selected-space second-moment ratio $\mathbb{E}[(m+u)^2]/\mathbb{E}[c_2^2]$ is below 1\% at all four sizes, including the within-block component $u$ (Fig.~\ref{fig:accounts}f; Supplementary Section~\ref{supp:selectedspace}). The loss decomposition links the predictive gains to complementary residual information: the prespecified XOR contrasts give positive simultaneous bounds for $T_r$ and $T_r-T_m$ at $n=10,14,16$ (Supplementary Fig.~\ref{supp:confirmation}). These findings motivate locating the unresolved components outside the selected space.

A separate paired Max3XOR study resolves the full response by Pauli weight, the number of nonidentity factors in a Pauli string. Each size uses 256 training and 512 evaluation instances, pairing A- and C-initialized preparations with the same C-centred continuation bank. We split the complement as $r=r_{\mathrm{low}}+r_{\mathrm{hi}}$, where $r_{\mathrm{low}}$ contains the unselected weight-zero to weight-two terms and $r_{\mathrm{hi}}=\sum_{w=3}^n r_w$ contains the higher-weight sectors. Applying Eq.~\eqref{eq:loss_identity} to this refined decomposition connects each sector to the actual scalar-prediction change (Methods~\ref{meth:xorweights}).

\begin{figure*}[tbp]
 \centering
 \manuscriptfigure[0.60\textheight]{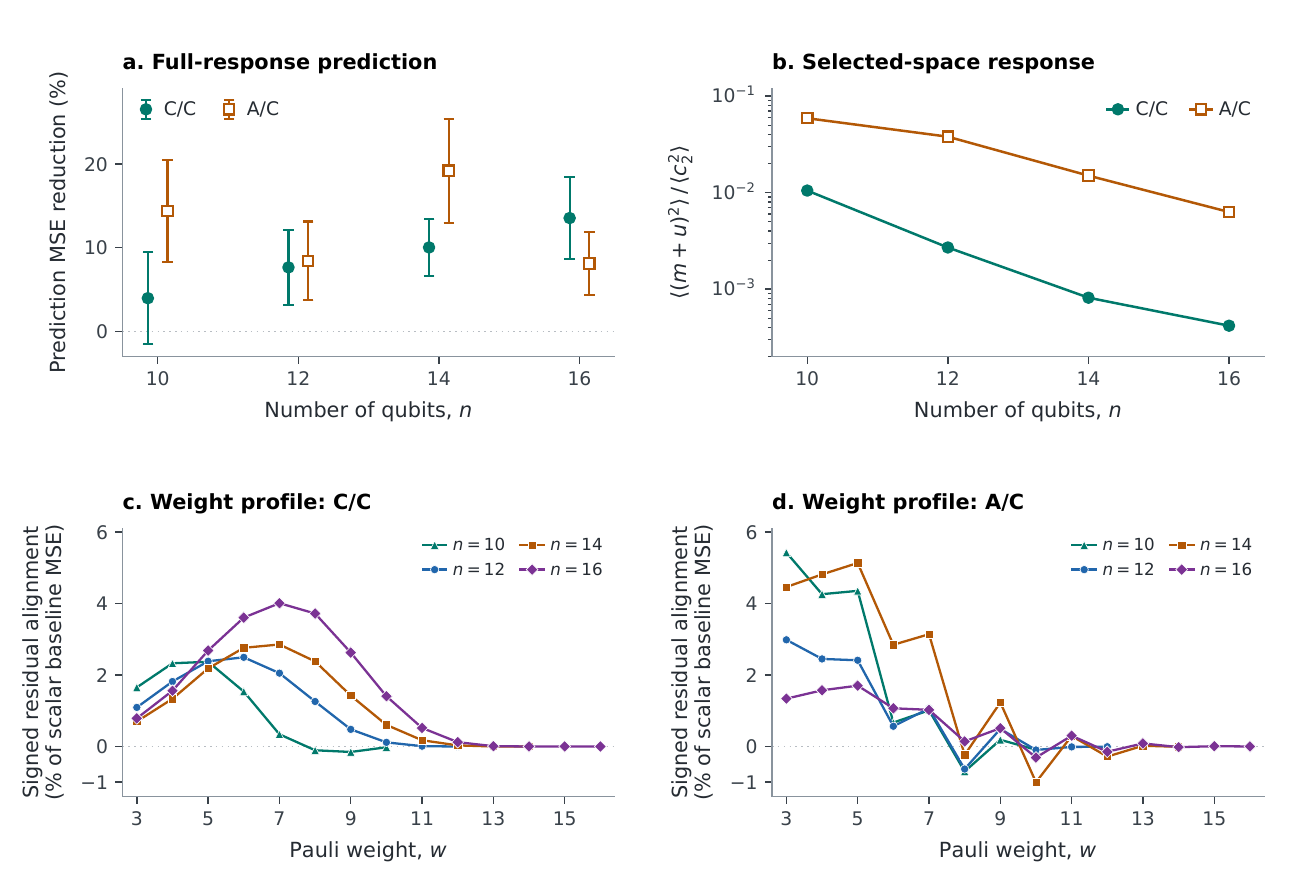}
 \caption{\textbf{Low-order summaries carry information about higher-weight response sectors.}
 The paired Max3XOR Pauli-weight study uses 256 training and 512 evaluation instances at each of $n=10,12,14,16$. AC and CC share the continuation bank and response operator within each instance. (a) Extra Trees/M3 scalar prediction-MSE reduction, with 95\% simultaneous intervals within the prespecified size-pair families (Methods~\ref{meth:xorweights}; Supplementary Table~\ref{supp:xorweighttable}). (b) The full selected-space second-moment ratio $\mathbb{E}[(m+u)^2]/\mathbb{E}[c_2^2]$, on a logarithmic scale; this ratio is not an additive response fraction. (c,d) Signed alignment $100T_w/R_B=200\mathbb{E}[(r_w-g_w)\delta]/R_B$ by Pauli weight in CC and AC, respectively, on the same scale. Panels (b--d) show descriptive point estimates.}
 \label{fig:xorweights}
\end{figure*}

The Pauli-weight cohorts reproduce the small selected-space response alongside useful M3 predictions (Fig.~\ref{fig:xorweights}a,b). In CC, the selected-space second-moment ratio is 1.051\% at $n=10$ and 0.0419\% at $n=16$, where the scalar MSE reduction is 13.54\% [8.62,18.46]. Across AC and CC at all four sizes, the higher-weight residual-alignment term $T_{\mathrm{hi}}=2\mathbb{E}[(r_{\mathrm{hi}}-g_{\mathrm{hi}})\delta]$ has positive simultaneous bounds. Seven of the eight scalar gains also have positive bounds; the smaller $n=10$ CC gain is 3.98\% [$-1.53$,9.48] (Supplementary Table~\ref{supp:xorweighttable}). Thus the extra information extends to higher-weight response sectors, beyond omitted one- and two-qubit terms.

The paired preparations distinguish the response operator from the state ensemble. Within each matched instance, AC and CC share $G$ and the entire continuation bank, hence the same averaged $K$ and its Pauli coefficients. Their weight-resolved residual-alignment profiles nevertheless differ (Fig.~\ref{fig:xorweights}c,d): CC has a broad positive profile over intermediate weights, whereas AC places more of its positive alignment at lower weights. At $n=16$, the largest point-estimate alignment occurs at weight seven for CC and weight five for AC. Thus the prepared-state ensemble and the residuals left by the baseline shape the predictive role of low-order observables beyond the operator's coefficient spectrum.

M3 also improves prediction of the combined higher-weight response $r_{\mathrm{hi}}$. On the same evaluation cohorts, adding M3 reduces this target's prediction MSE by 6.31--15.67\% in seven of eight conditions relative to its own baseline MSE, with positive exploratory pointwise 95\% bounds. Here the predictions are sums of the jointly learned sector outputs within each Extra Trees B or M3 model. At $n=10$ CC, the reduction is $-2.16\%$ [$-6.81$,2.27]. For the scalar target, Ridge also retains positive gains and higher-weight residual alignment in all eight conditions under exploratory pointwise intervals (Supplementary Section~\ref{supp:xorweights}).

\subsection{Predictive information in high-reward preparations}
\label{res:highreward}

Having established preparation dependence under a shared response operator, we test whether the summaries remain predictive in Max3SAT states optimized for high intermediate reward. The H procedure directly optimizes the intermediate state's mean satisfied-clause fraction using a recipe fixed on separate calibration instances (Methods~\ref{meth:highreward}). At each size, the paired preparation study crosses A and H preparations with A- and C-centred continuations, using a common baseline $B_{\mathrm{common}}$ that retains the state, circuit and response information available for both procedures. The mean satisfied-clause fraction rises from 0.804--0.808 under A to 0.954--0.956 under H (Fig.~\ref{fig:highreward}a). Preparation-quality and good-solution-probability diagnostics for this study are supplied in Supplementary Section~\ref{supp:highreward} and Source Data; Supplementary Section~\ref{supp:statequality} reports the corresponding audit of the exact-component cohorts.

With H preparation and the C continuation, M3 reduces Extra Trees prediction MSE by 6.89\%, 5.81\% and 8.39\% at $n=12,14,16$, respectively, with positive simultaneous bounds (Fig.~\ref{fig:highreward}b). At these three sizes, the C-minus-A continuation gain differences also have positive simultaneous bounds (Supplementary Table~\ref{supp:highrewardtable}). At $n=10$, the C-centred gain is smaller, 1.20\%, and its interval includes zero. The H/A primary intervals include zero at all four sizes. Thus high-reward states retain additional predictive information whose benefit depends on the continuation.

Higher preparation reward changes both the summaries' relative benefit and the scale of prediction errors. Within this paired common-baseline study, switching from A to H preparation at the C continuation reduces relative gains at $n=12,14,16$. The absolute H/C MSE reductions nevertheless retain about 69\%, 52\% and 86\% of their A/C values, respectively, as the baseline MSEs also change (Fig.~\ref{fig:highreward}c). The same preparation change makes the cancellation index more negative at $n=12,14$ but less negative at $n=16$. Positive direct and complementary residual alignments coexist in the H/C Extra Trees loss accounts at these three sizes (Supplementary Section~\ref{supp:highreward}). Together, these comparisons identify how the information retained by the summaries depends on preparation, continuation and the error left by the common baseline.

\begin{figure*}[tbp]
 \centering
 \manuscriptfigure[0.43\textheight]{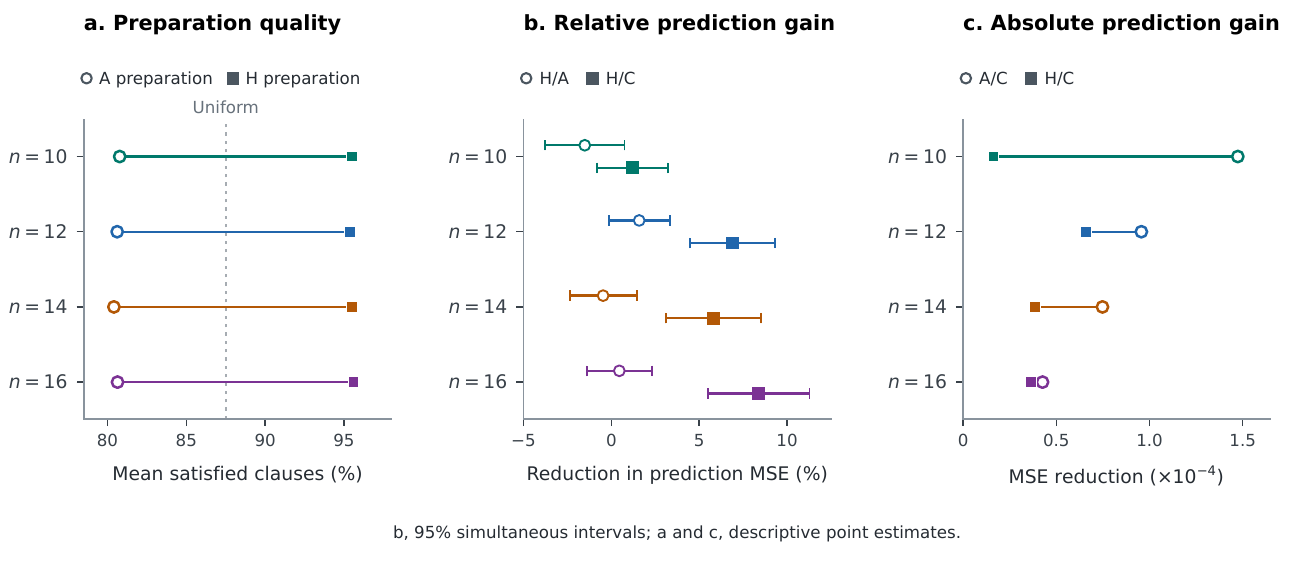}
 \caption{\textbf{High-reward preparation retains continuation-dependent predictive information.}
Each row shows one size in the paired Max3SAT preparation study. (a) Mean intermediate-state satisfied-clause fraction under A and H preparation; the dashed line marks the uniform-state value of $87.5\%$. (b) Extra Trees/M3 relative prediction-MSE reductions for H preparation followed by A- or C-centred continuations. Error bars show 95\% simultaneous intervals within each size's primary family (Methods~\ref{meth:highreward}). (c) Absolute MSE reductions $\Delta R=R_B-R_{\mathrm{M3}}$ for paired A/C and H/C conditions using the common baseline $B_{\mathrm{common}}$; these preparation-study cohorts are separate from those in Fig.~\ref{fig:gain}. Panels (a,c) are descriptive point estimates, with connecting lines pairing the two preparations. Each size uses 256 training and 512 independent evaluation instances, averaging repeated depths and generators within each evaluation instance. Supplementary Table~\ref{supp:highrewardtable} gives the paired gain and cancellation contrasts; full preparation-quality and absolute-loss diagnostics are supplied in Source Data.}
 \label{fig:highreward}
\end{figure*}

\subsection{Continuation controls connect observable usefulness to the response operator}
\label{res:continuation}

Having changed the prepared ensemble, we now hold the state fixed to isolate the role of subsequent dynamics. Matched-state Max3SAT discovery cohorts at $n=10,12,14,16$ share the A-initialized preparation, intervention generator and continuation perturbations within each instance and depth. Only the continuation centre changes, according to
\begin{equation}
 \boldsymbol\gamma(a)=(1-a)\boldsymbol\gamma_A+a\boldsymbol\gamma_C,
 \qquad
 \boldsymbol\beta(b)=(1-b)\boldsymbol\beta_A+b\boldsymbol\beta_C.
 \label{eq:angle_interpolation}
\end{equation}
Five diagonal points $a=b\in\{0,\tfrac14,\tfrac12,\tfrac34,1\}$ and the two cross conditions $(a,b)=(0,1),(1,0)$ separate joint changes from changes in one angle sector. The preparation uses the same A-initialized, instance-optimized recipe at every point, including the C-centred continuation endpoint.

Across the four sizes, condition-specific Extra Trees/M3 gains are small near A and the midpoint, while the C-centred and cost-angle-only changes give gains of 26.05--34.35\% (Fig.~\ref{fig:continuationphysics}; Supplementary Fig.~\ref{supp:continuationfigure}). Midpoint gains remain between $-0.25\%$ and 0.96\% even though the within-condition fluctuation cancellation defined above drops from strongly positive values at A to small negative values. Along the path, the mean X-block coefficient reverses sign, while its mean per-continuation block RMS at the midpoint falls to 30.84--32.12\% of the average endpoint RMS. The midpoint thus combines reduced cancellation with suppressed X-coefficient magnitude, while the full path distinguishes magnitude suppression from sign reversal (Supplementary Section~\ref{supp:transfer}).

An exact symmetry explains the coefficient sign reversal. For OR at all four sizes, the C centres on this path are close to a cost-angle reflection of A, with similar mixer angles (Methods~\ref{meth:calibrationcentres}). With $\mathcal Z=\prod_q Z_q$, real diagonal cost and intervention operators, and an X mixer, the response operator satisfies
\begin{equation}
 K(-\boldsymbol\gamma,\boldsymbol\beta)
 =\mathcal Z K(\boldsymbol\gamma,\boldsymbol\beta)^*\mathcal Z,
 \label{eq:gamma_reflection}
\end{equation}
where the star denotes entrywise complex conjugation. Pauli coefficients acquire the sign $(-1)^{N_X}$, where $N_X$ counts X factors: selected X coefficients are odd, XX and YY are even, and X coefficients vanish when every actual cost angle is zero. Reflecting the entire cost-angle vector, including its deviations, preserves coefficient magnitudes while reversing the X contribution at a fixed state. The A/C comparison shares deviations around approximately reflected centres and therefore does not exactly reflect each sampled circuit. The derivation and numerical checks appear in Supplementary Section~\ref{supp:symmetry}; the full response and predictive gain also depend on the state and complementary components.

Because the A-to-C cost-angle displacement exceeds its mixer-angle displacement, we also compare the two angle sectors using generator-normalized shifts around A and C. These shifts put continuation changes on a shared displacement scale (Methods~\ref{meth:matched}). At displacement magnitude $h=0.75$, opposite mixer shifts around A both increase relative and absolute prediction gains while changing within-condition fluctuation cancellation in opposite directions. Larger gains need not mean lower prediction errors: under the negative shift, both predictors' MSEs increase, with the baseline worsening more. Across the tested sizes, centres and displacement magnitudes, mixer shifts produce larger mean-squared response changes than cost-angle shifts, whereas the angle sector producing larger changes in predictive gain depends on the setting. Supplementary Fig.~\ref{supp:matchedfigure} and Table~\ref{supp:matchedcancellation} give the conditionwise maps and absolute losses.

\begin{figure*}[tbp]
 \centering
 \manuscriptfigure[0.61\textheight]{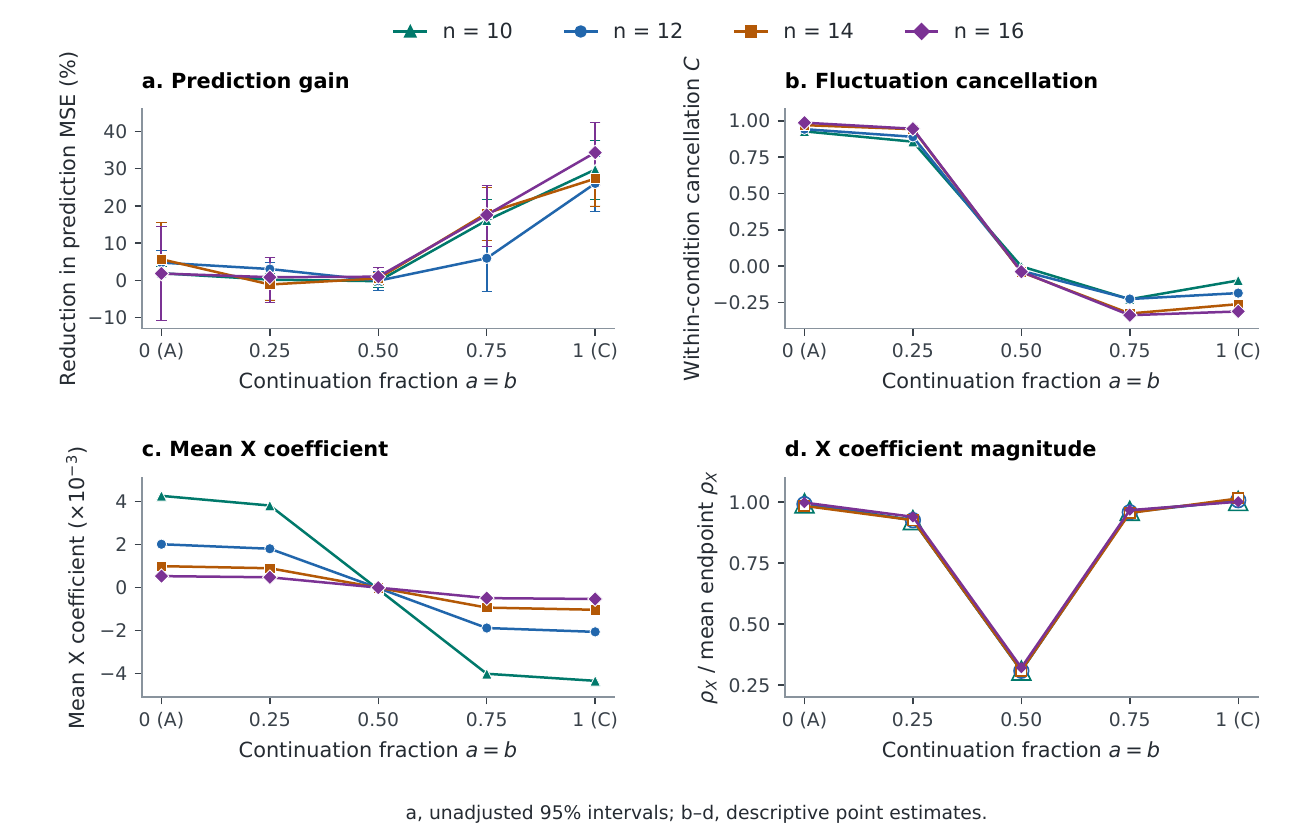}
 \caption{\textbf{Continuation dynamics reshape response structure and predictive value.}
 Matched-state Max3SAT discovery at $n=10,12,14,16$ along the five diagonal centres $a=b$ in Eq.~\eqref{eq:angle_interpolation}. The preparation, intervention generator and continuation deviations remain paired within each instance and depth. (a) Extra Trees/M3 relative prediction-MSE reduction; error bars are unadjusted 95\% exploratory intervals. (b) Cancellation after centering within depth--generator conditions. (c) Mean X-block response coefficient. (d) Mean per-continuation X-block coefficient RMS, divided by the mean of its A- and C-endpoint values at the same size. Panels (b--d) show descriptive point estimates. The midpoint combines weak predictive gain with reduced cancellation and suppressed X-coefficient magnitude; the coefficient changes sign along the path. Each size uses 256 training and 256 discovery-evaluation instances, with models fitted separately at each centre. Cost-only and mixer-only controls appear in Supplementary Fig.~\ref{supp:continuationfigure}a; generator-normalized controls at $n=10,12,14,16$ appear in Supplementary Fig.~\ref{supp:matchedfigure}.}
 \label{fig:continuationphysics}
\end{figure*}

\subsection{Frozen predictors retain their advantage at new angle combinations}
\label{res:transfer}

The continuation maps use models fitted separately at each centre. We next test whether the added information remains useful at unseen centres without refitting. We pooled the seven discovery-training conditions for 256 instances at each of $n=10,12,14,16$. Each size used four frozen predictors: B and M3 inputs for Ridge and Extra Trees. The models, sample sizes and comparisons were fixed before generating 512 fresh instances per size at four unseen interior centres, $(a,b)\in\{0.125,0.875\}^2$. Prepared states and continuation deviations remained paired across angle and generator conditions within each instance and depth. The same-size predictors were applied unchanged.

\begin{figure*}[tbp]
 \centering
 \manuscriptfigure[0.65\textheight]{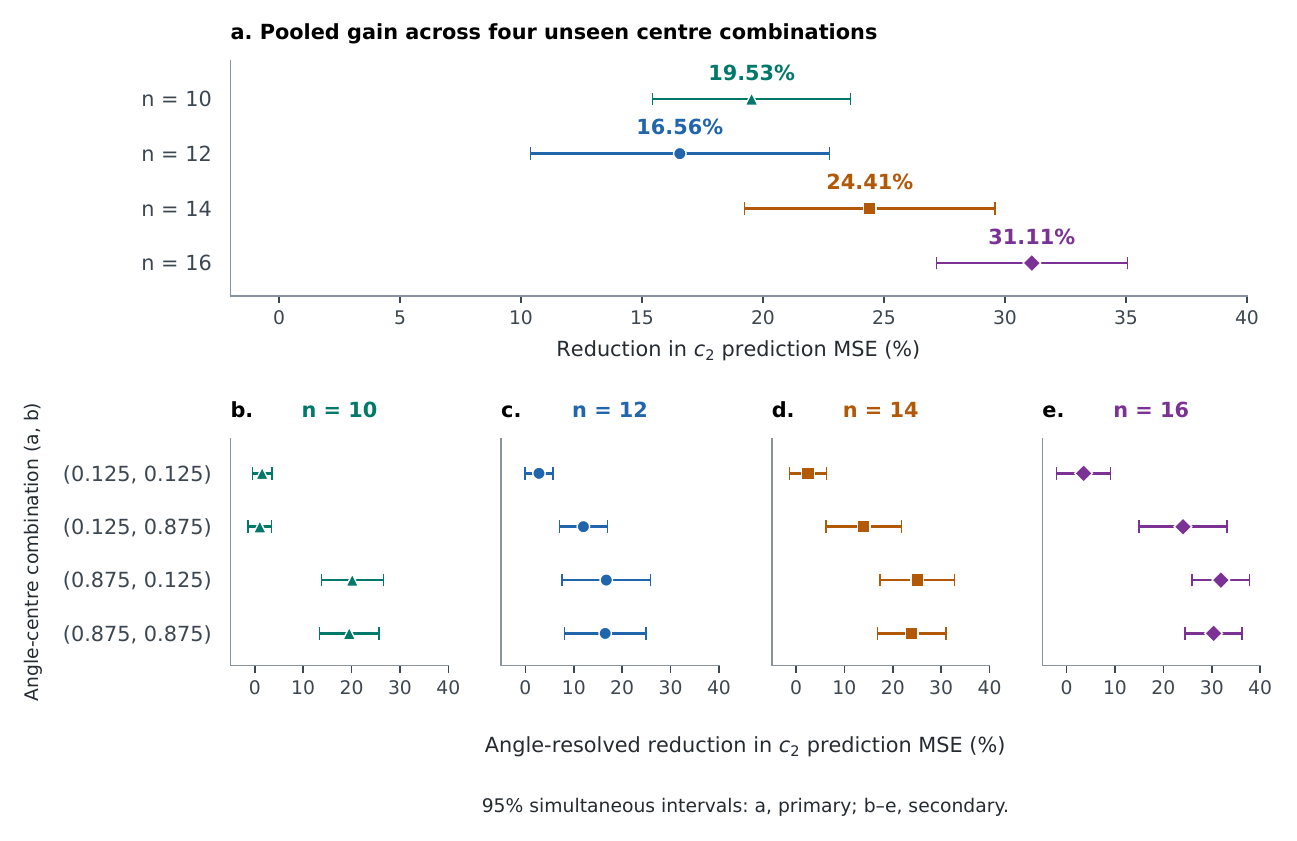}
 \caption{\textbf{Additional predictive information transfers to new continuation centres.}
 Extra Trees/M3 relative MSE reductions on 512 fresh Max3SAT instances per size, using predictors fixed before evaluation-instance generation. (a) Gains pooled over four unseen centre combinations, two depths and two intervention generators. (b--e) Conditionwise gains at $n=10,12,14,16$, respectively, for the interpolation fractions $(a,b)$ in Eq.~\eqref{eq:angle_interpolation}. Primary (a) and secondary (b--e) 95\% simultaneous intervals retain separate families for $n=10$, joint $n=12,14$, and $n=16$. Repeated conditions remain grouped by instance (Methods~\ref{meth:transfer}). All four pooled gains have positive bounds. Conditionwise results show variation within the tested interior of the A-to-C path.}
 \label{fig:transfer}
\end{figure*}

The prespecified Extra Trees/M3 gains are 19.53\%, 16.56\%, 24.41\% and 31.11\% at $n=10,12,14,16$, respectively, with positive simultaneous bounds at all four sizes (Fig.~\ref{fig:transfer}a). Ridge gains of 5.00--5.92\% also have positive simultaneous bounds. Absolute losses and additional prediction diagnostics appear in Supplementary Section~\ref{supp:transfer}. Thus the compact summaries retain predictive value on fresh problems at continuation centres not used for training.

Conditionwise evaluations locate the predictive advantage within the tested angle range. When both angle centres are nearer A, $(a,b)=(0.125,0.125)$, gains are 1.53--3.52\% across the four sizes, with every interval including zero (Fig.~\ref{fig:transfer}b--e). The primary gain pools raw losses before taking their ratio: the two C-nearer cost-angle conditions, $a=0.875$, contribute 98.45--99.93\% of baseline squared loss and therefore dominate this estimate. Conditionwise gains describe predictive usefulness in regions contributing little to the pooled loss (Supplementary Section~\ref{supp:transfer} and Table~\ref{supp:transferlosses}). Recomputing targets from each 32-continuation half while holding predictions fixed retains positive pooled gains.

\section{Discussion}

A compact observable's predictive role is determined jointly by the physical response, the prepared-state ensemble and the information already resolved by the baseline. The exact response and loss decompositions make this relationship explicit: they identify the unresolved components aligned with the realized prediction change, including opposing alignments that reduce the net gain. Max3XOR provides a particularly clear separation from direct response magnitude. Low-order summaries carry information about higher-weight response residuals even when their own selected space carries little response. Under the same response operator, paired preparations give different weight-resolved alignment profiles. The operator's coefficient spectrum alone therefore does not specify which residual information a state summary will reveal.

The matched-state continuation study supplies the complementary control: a fixed preparation supports different predictive benefits when the subsequent dynamics change. Reflection symmetry constrains direct coefficients, while generator-normalized shifts separate changes in cancellation from changes in predictive gain. Together with useful information in high-reward preparations, these controls show why state quality or a single response-magnitude diagnostic cannot replace a description of the state--continuation--baseline relationship. Frozen-model gains further demonstrate that the compact summaries retain predictive value at unseen centres within the trained range.

This account connects observable-based control~\cite{wauters2020,magann2022} and studies of state resources~\cite{caliz2025} to the design of predictive representations. Both Ridge and Extra Trees benefit from M3, with gains that depend on the learner and observable combination. A retrospective sensitivity analysis preserves the exact-component contrast directions at the point-estimate level when Ridge replaces the baseline component predictors while the Extra Trees scalar predictions remain fixed (Supplementary Section~\ref{supp:componentsensitivity}). Operational gains can arise from information additional to $B$, from making information already encoded in $B$ easier to learn, or from both. In each case, the response and loss decompositions identify the physical components associated with the realized improvement. This gives a concrete basis for comparing state summaries by the unresolved response information they provide.

The phase-sensitivity interpretation in Eq.~\eqref{eq:symmetric_response} suggests a direct use for prediction: estimating the leading effect of controlled phase fluctuations on success probability. A next step is to test observable selection or intervention screening under a matched experimental budget, comparing finite-shot predictions with direct sampling of perturbed circuits. Rules fixed on calibration problems can then be evaluated on fresh instances. Selecting a signed intervention additionally requires the linear response and validation at finite intervention strength. These tests would connect the physical diagnosis established here to the use of compact state information in quantum control~\cite{magann2022}.

\section{Methods}
\label{meth:main}

The exact-component, Pauli-weight-resolved Max3XOR, high-reward preparation, seven-centre continuation, generator-normalized control and prospective transfer studies cover $n=10,12,14,16$ and share the physical definitions below. Component-predictor sensitivity uses the saved exact-component cohorts at all four sizes. Calibration schedules remain fixed within each size (Section~\ref{meth:calibrationcentres}), so cross-size comparisons assess recurrence under these calibrations rather than a scaling law with a common calibration history.

\subsection{Problem ensembles and response target}
\label{meth:problems}
The exact-component studies used signed three-variable constraint problems with $n\in\{10,12,14,16\}$ and $2n$ clauses. Distinct variable triples were sampled without replacement, and each literal received a binary sign. The same labelled triples and signs defined paired Max3SAT (OR) and Max3XOR (XOR) instances. An OR clause was satisfied when at least one literal was true; an XOR clause was satisfied when an odd number were true. We use ``instance'' for this shared signed-clause specification. Duplicate specifications were excluded by their canonical labelled representation, including signs, rather than by graph isomorphism.

Let $R(z)$ count the satisfied clauses of bitstring $z$. Circuit optimization maximized $\langle R\rangle/(2n)$, evaluated after the full circuit for A- and C-initialized preparation and after the preparation segment for H (Sections~\ref{meth:circuits} and~\ref{meth:highreward}). The diagonal cost Hamiltonian had entries $R(z)-n$. Exact enumeration gave the attained minimum and maximum rewards, $R_{\min}$ and $R_{\max}$. We defined
\begin{equation}
 \Pi_{\mathrm{good}}=\sum_{z:\,R(z)\ge R_{\max}-0.05(R_{\max}-R_{\min})-10^{-10}}
 |z\rangle\langle z|.
 \label{meth:goodset}
\end{equation}
The threshold selects the upper five per cent of the attained reward range, not five per cent of the bitstrings. Its membership can include optimal and near-optimal solutions.

We evaluated $c_{2,\theta}=P_\theta''(0)/2$ in Eq.~\eqref{eq:response_series} directly, using the projected vectors $q_0=\Pi_{\mathrm{good}}V_\theta\psi$, $q_1=\Pi_{\mathrm{good}}V_\theta G\psi$, and $q_2=\Pi_{\mathrm{good}}V_\theta G^2\psi$:
\begin{equation}
 c_{2,\theta}=\langle q_1|q_1\rangle-\operatorname{Re}\langle q_0|q_2\rangle.
 \label{meth:directc2}
\end{equation}
Averaging over 64 sampled continuations gives the target $c_2$. The labels therefore use state-vector propagation without finite-difference differentiation.

\subsection{Preparation and continuation circuits}
\label{meth:circuits}
QAOA circuits alternated cost evolution and the mixer $\exp(-i\beta\sum_qX_q)$, starting from $|+\rangle^{\otimes n}$. For total depth $p\in\{4,8\}$, the first $p/2$ layers prepared $|\psi\rangle$, and the continuation contained $p/2$ layers. We distinguished analytic (A) and calibrated (C) schedules; the high-reward preparation H is defined in Section~\ref{meth:highreward}. In a two-letter condition such as AC, the first letter denotes the preparation schedule and the second the continuation centre.

For analytic preparation, $x_j=(j+1/2)/p$, $j=0,\ldots,p-1$, defined the initial angles
\begin{equation}
 \gamma_j=-(0.2+0.8x_j),\qquad \beta_j=0.05+0.5(1-x_j).
 \label{meth:analytic}
\end{equation}
Independent Gaussian perturbations with standard deviation $0.15$ were added to the angles. Calibrated preparation instead used a previously fixed calibration schedule with the same perturbation scale. These A- and C-initialized preparations performed 40 simultaneous perturbation stochastic approximation (SPSA) updates~\cite{spall1992,spall1998} on the full circuit. At update $k=0,\ldots,39$, the perturbation and learning scales were $0.10/(k+1)^{0.101}$ and $0.12/(k+1)^{0.602}$, with Rademacher perturbation directions. The best objective value among the initial and updated iterates selected the final angle vector; its first half defined the intermediate state. Calibration schedules were fixed inputs to the confirmation, not refitted on its evaluation instances. Their values and size-dependent calibration designs are given in Section~\ref{meth:calibrationcentres}.

Continuations were centred on the final half of the analytic or calibrated schedule, with independent Gaussian angle perturbations of standard deviation $0.15$. Each instance--depth pair had its own bank of 64 perturbation draws, shared across its matched conditions and intervention generators. The exact-component confirmation comprised OR AA, OR AC, phase-transformed OR AC, and XOR CC.

Two intervention families were used. The first generator was a linear combination of all single-site $Z_q$ operators; the second combined all pairs $Z_qZ_r$. Coefficients were independent standard normal draws, normalized to unit Euclidean norm. Generator normalization did not depend on the prepared state.

The phase-transformed OR AC condition applied
\begin{equation}
\begin{aligned}
D&=\exp\!\left(-i\sum_{q=0}^{n-1}a_qZ_q\right),\\
 a_q&=\frac{\pi}{8}(-1)^q\frac{1+(q\bmod 3)}{3}
\end{aligned}
 \label{meth:phasecontrol}
\end{equation}
to the intermediate state before the intervention. This transformation preserves computational-basis probabilities and diagonal cost statistics. All input features were recomputed on the transformed state, so mixer-response features could change. Identity and transformed conditions shared the original instance, preparation, intervention generator, and continuation draws.

\subsection{Observable summaries and exact components}
\label{meth:components}
For the individual Pauli operators entering the summaries in Eq.~\eqref{eq:three_summaries}, we use $\alpha_P=2^{-n}\operatorname{Tr}(K_\theta P)$ and $e_P=\langle\psi|P|\psi\rangle$, with $K_\theta$ defined in Eq.~\eqref{eq:response_operator}.

Since $G|z\rangle=\lambda_z|z\rangle$, the computational-basis entries obey
\begin{equation*}
 (K_\theta)_{zz'}=-\tfrac12(\lambda_z-\lambda_{z'})^2(A_\theta)_{zz'}.
\end{equation*}
Hence $(K_\theta)_{zz}=0$ and $2^{-n}\operatorname{Tr}(K_\theta Z_S)=0$ for every product $Z_S=\prod_{q\in S}Z_q$. This vanishing direct coefficient does not prevent diagonal state observables from predicting $c_2$ through correlations in the prepared ensemble. The baseline also contains mixer-probe qualities, which need not be diagonal observables of the intermediate state.

This uses the standard Heisenberg-picture response expansion\cite{hadfield2023}. If $\bar\alpha_k$ is the mean coefficient in block $k\in\{X,XX,YY\}$, the decomposition is
\begin{equation}
\begin{aligned}
m&=\sum_k\bar\alpha_k\sum_{P\in k}e_P,\\
 u&=\sum_k\sum_{P\in k}(\alpha_P-\bar\alpha_k)e_P,\\
 r&=c_2-\sum_{P\in\mathcal P}\alpha_Pe_P,\\
 c_2&=m+u+r.
\end{aligned}
 \label{meth:decomposition}
\end{equation}
The definitions apply separately to each continuation and then to their averages. The term $m$ is the block-average contribution associated with the three summaries; $u$ retains coefficient variation within the selected blocks; $r$ contains the contribution outside the selected space. Operator coefficients were evaluated over the full projector rank, using its complement and the corresponding sign relation when that reduced cost. No rank truncation was used at any of the four sizes, $n=10,12,14,16$. Exact coefficients and component targets were used for diagnosis and supervised labels, never as scalar-prediction inputs.

\subsection{Baseline features and supervised models}
\label{meth:learning}
The baseline $B$ included all single-site $Z$ and pairwise $ZZ$ expectations; cost mean and variance; normalized computational-basis Shannon entropy; a 40-bin reward histogram; mixer-probe qualities at angles $\pm0.15$ and their differences; circuit angles, size, depth, and preparation objective; signed-clause degrees and pair counts; known generator coefficients and its state mean and root-mean-square value; the continuation centre; and known phase-control descriptors. Control descriptors included a fixed six-category encoding and the single-site and pairwise phase coefficients, supplied equally to both predictors. Thus $B$ includes mixer-response information. Representing this information as a single input vector gives baseline input dimensions of 352, 446, 556 and 682 for $n=10,12,14,16$, respectively. We compare the predictive performance of this baseline with that of a model augmented by the three observable summaries.

Writing $q_z=|\langle z|\psi\rangle|^2$ and $r_z=R(z)/(2n)$, histogram bin $k\in\{0,\ldots,39\}$ contains $\sum_{z:\,\min(\lfloor40r_z\rfloor,39)=k}q_z$; the normalized entropy is $-\sum_{q_z>0}q_z\log_2q_z/n$. The mixer-probe qualities are the expected normalized reward after applying $\exp(\mp i0.15\sum_qX_q)$ to the current state, and their difference feature is $(Q_+-Q_-)/0.30$.

We compared $B$ with four augmentations: X, XX, M2, and M3, adding respectively $s_X$, $s_{XX}$, $(s_X,s_{XX})$, and $(s_X,s_{XX},s_{YY})$. Following earlier development studies, M3 was fixed as the primary representation before the independent evaluations. Comparisons with X, XX and M2 assess the predictive value of individual summaries and the additional predictive value obtained by combining them. The feature allowlist excluded post-intervention good probabilities, response and component labels, exact operator coefficients, and Monte Carlo error labels. Current good-set probability was not a separate feature, but the histogram resolves integer reward levels at these sizes and determines it for a known threshold.

Ridge regression~\cite{hoerl1970} used penalties $\{0.1,10,1000\}$ and input standardization fitted only on each training fold. Extra Trees\cite{geurtsextratrees2006} used 128 trees and minimum leaf sizes $\{1,4,12\}$. Both learners were implemented in scikit-learn~\cite{pedregosa2011}. Training-fold target means and standard deviations set the target scaling for both learners; predictions were returned to the original units. Scalar $c_2$ and joint three-output $(m,u,r)$ models were trained separately. Hyperparameters minimized the summed standardized squared error in three-fold cross-validation grouped by instance, so all four depth--generator observations of an instance remained in one fold. The selected model was refitted on all training instances. Fold assignment and model random-number streams were shared across matched conditions and feature variants. All predictors were fixed before evaluation panels were assembled and accessed.

\subsection{Sampling and statistical analysis}
\label{meth:statistics}
Each exact-component study used 256 training and 512 evaluation instances per size. Training, evaluation and numerical-validation cohorts were disjoint, and pilot instances were excluded from the reported results. New signed-clause specifications and identifiers were screened for overlap with the other cohorts, reserved calibration problems and previous studies; intentionally reused cohorts are identified separately. Each condition had four observations per instance, corresponding to two depths and two generators. A condition panel therefore contained 2,048 evaluation rows but 512 independent instances.

For the risks in Eq.~\eqref{eq:gain}, we first averaged the four squared prediction errors within each instance and then weighted instances equally. The prediction-loss analysis retained all five signed terms in Eq.~\eqref{eq:loss_identity}; physical cancellation used Eq.~\eqref{eq:cancellation}.

The primary model and representation were Extra Trees and M3. Directional contrasts were fixed before generating the new instances in three separate primary families: seven at $n=10$, eleven jointly at $n=12,14$, and seven at $n=16$ (Supplementary Fig.~\ref{supp:confirmation} and Section~\ref{supp:primary}). The corresponding secondary families contained 1,408, 2,816 and 1,408 metrics. We used 4,000 paired instance-level bootstrap replicates~\cite{efron1979}. All conditions and predictors at a given size shared multinomial resampling weights; sizes used independent random-number streams. For contrast $j$, let $s_j$ be its bootstrap standard deviation. The 95th percentile $q$ of the bootstrap maximum $\max_j|\widehat\theta_j^*-\widehat\theta_j|/s_j$ gave intervals $\widehat\theta_j\pm q s_j$~\cite{montielolea2019}, separately within each stated family. Ratios, including baseline-MSE denominators, were recomputed within each replicate. Each resampled instance retains its realized continuation banks and repeated conditions; training data, fitted predictors and calibration schedules remain fixed. The intervals describe instance-bundle variation under this finite-bank design, without separately estimating bank replacement for fixed instances, refitting variation or uncertainty for an infinite-continuation target. The nominal 95\% simultaneous coverage is approximate and family-specific. Supplementary analyses distinguish the exact-component cohorts from the scalar, learning-size and calibration-transfer evaluations.

Retrospective diagnostics use the saved exact-component cohorts at all four sizes. Half-bank analyses recompute targets from each 32-continuation half while retaining the fitted predictions and grouping all repeats by instance; their contrast families remain separate by study (Supplementary Section~\ref{supp:bank}). Pooled and depth--generator-specific centering are descriptive analyses of the component targets (Supplementary Section~\ref{supp:centering}). The Z2 control retains the original training and evaluation splits, fits the new diagonal-summary models with training-only selection, and fixes them before re-evaluating the previously examined cohorts. Its two eight-contrast main families and two 88-metric secondary families remain separate for the joint $n=12,14$ and joint $n=10,16$ comparisons (Supplementary Section~\ref{supp:diagonalcontrol}). The reward audit replays prepared states and the same 64 continuations at $d=0$, counting the two generator labels only once and reporting descriptive means over 512 instances per size (Supplementary Section~\ref{supp:statequality}).

\subsection{Sensitivity to the component predictor}
\label{meth:componentsensitivity}
We re-evaluated the prediction-loss account on the saved exact-component cohorts at $n=10,12,14,16$, keeping the Extra Trees scalar predictions $f_B$, $f_{\mathrm{M3}}$ and $\delta=f_{\mathrm{M3}}-f_B$ fixed. Only the baseline component predictions $(g_m,g_u,g_r)$ were replaced by the previously fitted Ridge predictions. Both learners used the original training split and training-only model selection. We recomputed all five terms in Eq.~\eqref{eq:loss_identity}, including the correction for the separately fitted scalar and component predictors, together with component MSE and MSE divided by target variance. This isolates changes in the component account while preserving the measured scalar gain and physical targets.

The retrospective analysis evaluated the 21 predictor-dependent contrasts from the original physical studies across all four sizes. The four physical cancellation contrasts are invariant to this replacement and were retained as checks. One 21-contrast family assessed the Ridge-component values; a separate 21-contrast family assessed their paired differences from the Extra Trees-component values. The 4,000 whole-instance bootstrap replicates and maximum-standardized intervals followed Section~\ref{meth:statistics}, pairing conditions and component predictors within each size. The original confirmation families were retained, and no models were refitted or evaluation instances selected in this sensitivity analysis.

\subsection{Pauli-weight-resolved Max3XOR study}
\label{meth:xorweights}
Two new studies covered $n=10,12$ and $n=14,16$, each with 256 training and 512 evaluation Max3XOR instances per size. Within each study, canonical signed-instance specifications were distinct across training, evaluation and the two excluded benchmark instances per size. Preparations A and C followed Section~\ref{meth:circuits}, paired with the native fixed C continuation centre at each size. The AC and CC conditions shared the instance, normalized Z/ZZ generator and 64 continuation draws at each depth $p=4,8$, and therefore shared each response operator $K_\theta$ and its bank average. Each preparation supplied four depth--generator rows per instance.

We refined the exact decomposition to $c_2=m+u+r_{\mathrm{low}}+\sum_{w=3}^{n}r_w$. Here Pauli weight $w$ counts nonidentity sites, $r_{\mathrm{low}}$ contains all weight-zero, -one and -two terms outside the selected space $\mathcal P$, and $r_w$ sums $\alpha_P e_P$ over weight-$w$ strings. All $4^n$ Pauli terms and the full good-set projector rank were retained, using a dense tensor transform at $n=10,12$ and an exact streamed expansion at $n=14,16$. The signed sector contributions include their cross moments when reconstructing the total response second moment. We write $r_{\mathrm{hi}}=\sum_{w=3}^{n}r_w$. Its residual-alignment term is $T_{\mathrm{hi}}=2\mathbb E[(r_{\mathrm{hi}}-g_{\mathrm{hi}})\delta]$, where $g_{\mathrm{hi}}=\sum_{w=3}^{n}g_w$ sums baseline component predictions and $\delta=f_{\mathrm{M3}}-f_B$. For this refined account, $g_r=g_{\mathrm{low}}+\sum_{w=3}^{n}g_w$ and $d_B=f_B-g_m-g_u-g_r$ use the newly fitted sector outputs. The low-weight and individual weight alignments then sum to $T_r$, preserving Eq.~\eqref{eq:loss_identity} with its corresponding correction and prediction-change penalty.

The baseline in Section~\ref{meth:learning} retained all preparation and continuation context, with phase-control descriptors set to their identity values; M3 added the same three summaries. Extra Trees used 256 trees and minimum leaf sizes $\{1,4,12\}$; Ridge retained the penalties in Section~\ref{meth:learning}. Scalar and joint sector-output models were fitted separately for each size and preparation. Instance-grouped three-fold cross-validation minimized mean training-fold-standardized squared error across outputs, following the fold-local preprocessing and model-fitting protocol in Section~\ref{meth:learning}. Within each condition, learner and target, B and M3 shared folds and estimator random-number streams. Exact sector targets and operator coefficients served only as labels and diagnostics.

Losses and signed terms first averaged the four rows within each instance, then weighted instances equally. The 4,000 paired whole-instance bootstrap replicates followed Section~\ref{meth:statistics}, sharing weights across preparations and learners within a size and using independent streams across sizes. Each study retained an eight-effect primary family: four Extra Trees/M3 scalar gains and four high-weight residual-alignment terms for its size pair. Approximate 95\% simultaneous intervals used the maximum-standardized construction, recomputing baseline-MSE denominators in every replicate. All other intervals were exploratory pointwise percentile intervals. Supplementary Section~\ref{supp:xorweights} reports the primary results and diagnostic Source Data.

\subsection{High-reward preparation and matched response comparison}
\label{meth:highreward}
Preparation H directly maximized the intermediate-state satisfied-clause fraction $Q_\psi=\langle\psi|R|\psi\rangle/(2n)$ over the $p/2$ prefix layers. Three initializations used zero, A-prefix and C-prefix angles, with independent Gaussian perturbations of standard deviation $0.03$, $0.15$ and $0.15$, respectively. Each run used the SPSA scales in Section~\ref{meth:circuits} and retained the best initial or updated iterate. The highest-$Q_\psi$ candidate among the three runs and the exact uniform state was selected. A separate calibration on 32 $n=12$ instances compared 40 and 200 updates per start along the same trajectories. We selected the shorter budget meeting mean $Q_\psi\ge0.895$ and at least 80\% of instances above the uniform value $0.875$, at both depths. The selected 40-update recipe was fixed and independently checked on eight $n=14$ instances before the response studies, then applied without retuning at all four sizes; selection used $Q_\psi$ alone.

The response studies used 256 fresh training and 512 fresh evaluation Max3SAT instances at each of $n=10,12,14,16$, disjoint from calibration and earlier cohorts. Each instance paired preparations A and H with continuation centres A and C, retaining $p=4,8$, normalized Z/ZZ generators and 64 matched continuation deviations. Every instance was retained. To compare the two preparations on common inputs, $B_{\mathrm{common}}$ removed the same 35 full-circuit context entries from both: the full angle vector, its initialization centre, and the initial and best full-circuit objectives and their difference. State cost, probability-distribution and mixer-probe features, problem and generator descriptors, and the actual prefix and continuation angles remained. The common baseline had 317, 411, 521 and 647 inputs at $n=10,12,14,16$; M3 added three summaries. Scalar and component models were fitted separately in each cell using the Ridge and 128-tree Extra Trees grids and the fitting protocol in Section~\ref{meth:learning}. This evaluates newly trained predictors in the high-reward regime under different preparation objectives and budgets.

At each size, the two primary contrasts were the Extra Trees/M3 relative MSE gains for H followed by A and C. A separate two-contrast family tested the C-minus-A gain difference and the A-minus-C physical cancellation-index difference within H. The 4,000 paired whole-instance bootstrap replicates and maximum-standardized intervals followed Section~\ref{meth:statistics}, separately for each family and size, keeping all depth--generator observations and preparation--continuation cells paired. Absolute losses, preparation changes, signed loss terms and Ridge results were descriptive diagnostics, with pointwise intervals where reported.

\subsection{Controlled continuation discovery and prospective transfer}
\label{meth:transfer}
\paragraph{Discovery design.}
The continuation studies at $n=10,12,14,16$ used Max3SAT OR, the A-initialized preparation with 40 optimization updates, $p=4,8$, and the same normalized Z/ZZ intervention families. Each discovery size used 256 new training and 256 disjoint discovery-evaluation instances. The five joint interpolation points and two cross conditions in Eq.~\eqref{eq:angle_interpolation} shared the prepared state, generator and 64 Gaussian continuation deviations for each instance and depth. Only the continuation half of the centre was changed; unused prefix-centre entries in the input descriptors remained analytic. The input protocol followed Section~\ref{meth:learning}, with centre descriptors supplied equally to both models.

Discovery predictors were fitted separately at each of the seven conditions for B, X, XX, YY and M3. Ridge followed the fitting protocol in Section~\ref{meth:learning}. Extra Trees used 256 trees and a fixed minimum leaf size of four. The conditionwise results used 1,000 paired instance-bootstrap replicates and exploratory percentile intervals. They informed subsequent hypotheses and were not included in a prospective simultaneous family.

\paragraph{Frozen pooled predictors.}
For prospective evaluation, we combined only the discovery-training rows from all seven conditions: $256\times7\times2\times2=7,168$ rows per size. All repeats of an instance remained grouped in Ridge cross-validation. Extra Trees retained 256 trees, minimum leaf size four and all features available at each split. Scalar target standardization used training values only. B and M3 models for both learners were fitted separately at each size and fixed before the prospective evaluation was generated. These pooled predictors are distinct from the condition-specific discovery predictors.

\paragraph{Fresh evaluation cohort.}
The evaluation used 512 fresh instances per size and $(a,b)\in\{0.125,0.875\}^2$, preserving paired preparations and continuation deviations across conditions. Each evaluation instance contributed 16 rows, from four angle-centre combinations, two depths and two generators, giving 8,192 rows per size and 2,048 independent transfer instances across the four sizes. The target was the direct statevector response averaged over 64 continuations. This stage evaluated response prediction, with the exact-component diagnosis supplied by the preceding studies.

\paragraph{Estimands and inference.}
For the primary comparisons we first averaged raw squared losses equally over each instance's 16 repeated conditions and then over instances, before forming $1-R_{\mathrm{M3}}/R_B$. This estimand differs from the average of the four conditionwise percentage reductions. The overall Extra Trees gain at each size was the prespecified primary contrast. The three studies retain separate primary and secondary families: one primary and 23 secondary contrasts at $n=10$, two primary and 46 secondary contrasts jointly at $n=12,14$, and one primary and 23 secondary contrasts at $n=16$. Secondary contrasts covered Ridge gains, conditionwise and depth--generator gains, cost-angle fraction contrasts and fixed-prediction half-bank sensitivity. The cost-angle contrast subtracts the gain at $a=0.125$ from that at $a=0.875$, each computed by pooling raw losses over the two mixer-angle fractions, depths and generators.

The 4,000 paired instance-level bootstrap replicates and maximum-standardized intervals followed Section~\ref{meth:statistics}, with separate primary and secondary families and the same fixed-training, finite-bank interpretation. Each size uses its own fixed predictors, defining an unseen-centre interpolation test within size.

\subsection{Generator-normalized continuation contrasts}
\label{meth:matched}
The normalized-displacement map used a separate Max3SAT OR cohort with 256 training and 256 disjoint discovery-evaluation instances at each of $n=10,12,14,16$. These cohorts were disjoint from the exact-component and seven-centre studies. The A-initialized preparation with 40 optimization updates, depths $p=4,8$, normalized Z/ZZ generators and bank of 64 continuation deviations were retained. Each instance and depth shared its prepared state and deviations across all 18 conditions; A and C specify the continuation centres.

Let $L=p/2$ and $\boldsymbol d=(1,\ldots,1)/\sqrt L$ on the continuation layers. For the actual cost Hamiltonian $H_C$, define its uniform Hilbert--Schmidt variance by
\begin{equation}
 \sigma_C^2=2^{-n}\operatorname{Tr}(H_C^2)
 -\bigl[2^{-n}\operatorname{Tr}(H_C)\bigr]^2.
 \label{meth:cost_scale}
\end{equation}
The corresponding standard deviation of the mixer $\sum_qX_q$ is $\sqrt n$. We changed one angle sector at a time,
\begin{equation}
\begin{gathered}
\Delta\boldsymbol\gamma=\frac{s h}{\sigma_C}\boldsymbol d,
 \quad \text{or}\quad
 \Delta\boldsymbol\beta=\frac{s h}{\sqrt n}\boldsymbol d,
 \\ s\in\{-1,+1\},\quad h\in\{0.25,0.75\}.
\end{gathered}
 \label{meth:matched_shift}
\end{equation}
The squared gatewise generator-weighted displacement is then
$\sigma_C^2\|\Delta\boldsymbol\gamma\|^2+n\|\Delta\boldsymbol\beta\|^2=h^2$.
Each centre contributed its baseline and eight signed shifts. Throughout, equal length refers only to this gatewise generator-weighted displacement. The metric uses only the known generators; it does not equate full-unitary, final-state or quantum-Fisher distances. The uniform layer direction also differs from the original A-to-C interpolation direction.

Condition-specific B, X, XX, YY and M3 fits followed Sections~\ref{meth:learning} and~\ref{meth:transfer}. Extra Trees retained 256 trees and minimum leaf size four. Each condition had 1,024 evaluation rows from 256 independent instances. We used 1,000 paired whole-instance bootstrap draws under the fixed-model, finite-bank design in Section~\ref{meth:statistics}, forming unadjusted 95\% percentile intervals. The response-displacement and gain-change summaries in Supplementary Section~\ref{supp:matched} and its Source Data are descriptive contrasts without inferential intervals.

\subsection{Evaluation- and training-sample robustness}
\label{meth:n10}

A separate $n=10$ follow-up assessed evaluation- and training-sample variation, with its design and primary contrasts fixed before generating the new data. The scalar B/M3 predictors from the exact-component study were evaluated on 1,024 new instances. Five disjoint 256-instance training cohorts supplied additional scalar predictors under the same fitting procedure, hyperparameter candidates and model seeds. All new predictors were fixed before the common evaluation panels were opened. The two primary contrasts were the original-model and five-cohort mean-risk Extra Trees/M3 gains for XOR CC; 54 other contrasts formed a separate secondary family. The mean-risk gain averages squared losses over the five fitted models before taking the risk ratio; it does not average predictions or percentage gains.

The sample-robustness study used 4,000 paired whole-instance bootstrap replicates with the simultaneous-interval construction of Section~\ref{meth:statistics}, separately within its stated families. Primary intervals condition on the fitted models, including all five training realizations. A two-level resampling of training cohorts and evaluation instances was reported separately as a five-cohort sensitivity analysis. The new 1,024-instance evaluation remained distinct from the 512-instance exact-component panel; their pooled result was descriptive. Supplementary Section~\ref{supp:n10robustness} gives the robustness results.

\subsection{Fixed calibration centres and circuit conventions}
\label{meth:calibrationcentres}

Full-precision A and original C vectors for $n=10,12,14,16$ are supplied in the accompanying \texttt{calibration\_centres.json}. C depends on size, predicate and depth. A is identical across sizes and predicates at a fixed depth and is defined exactly by Eq.~\eqref{meth:analytic}.

All angles are in radians. A full vector stores $(\gamma_0,\ldots,\gamma_{p-1},\beta_0,\ldots,\beta_{p-1})$. Layer $j$ applies $\exp(-i\gamma_jH_C)$ followed by $\exp(-i\beta_j\sum_qX_q)$, with $H_C(z)=R(z)-n$. Preparation uses indices $0\le j<p/2$ of the instance-optimized full vector; the continuation uses indices $p/2\le j<p$ of the fixed centre plus its sampled perturbations. The 40 preparation updates therefore do not redefine the continuation centre. Both the A--C interpolation and generator-normalized displacements use these same fixed OR vectors.

For the original OR centres at $n=10,12,14,16$ and $p=4,8$, the largest entrywise differences are $|\gamma_C+\gamma_A|=0.040016$ and $|\beta_C-\beta_A|=0.055010$ radians. Thus C is close to the cost-angle reflection of the A centre. The coefficient reflection in Eq.~\eqref{eq:gamma_reflection} reverses actual cost angles including their perturbations; sharing the same perturbations around A and C does not produce such an exact reflected pair.

The original C schedules for $n=10,12$ were calibrated on 64 reserved instances per size with 100 SPSA updates per restart; those for $n=14,16$ used four reserved instances per size and 80 updates per restart. For each predicate and depth, the objective was the mean normalized full-circuit reward over that size's calibration instances. Two restarts initialized $\gamma_j=\pm(0.2+0.8x_j)$ and $\beta_j=0.05+0.5(1-x_j)$, using the SPSA scales stated in Section~\ref{meth:circuits}. Each restart retained the best initial or updated iterate; the restart with the larger retained objective supplied C. The calibration instances were shared across predicates and depths within each size and were reserved from training and evaluation. The schedules remained fixed throughout the confirmation studies, so size comparisons retain distinct calibration anchors.

\paragraph{Matched-budget comparison.}
The matched-budget robustness study at $n=14,16$ used three disjoint calibration cohorts of 64 instances per size that supplied new C schedules for both predicates and depths, with 100 updates per restart and the same two initializations and reward objective. All three schedule sets were retained. A new cohort of 256 training and 512 evaluation instances per size was shared by the original and three new C sets, disjoint from calibration and earlier cohorts. OR AC shared its A-prepared states across C sets; XOR CC preparations were recomputed from each C with paired initialization and optimization randomness. Depths, generators and 64 continuation deviations remained matched. Scalar B/M3 models followed Section~\ref{meth:learning}, including the 128-tree Extra Trees grid; all fitted models were fixed before evaluation.

Each instance supplies four depth--generator observations, whose squared errors are averaged before computing the instance-weighted risks.

We used 4,000 paired whole-instance bootstrap replicates and the maximum-standardized intervals of Section~\ref{meth:statistics}. Resampling weights were shared across schedules, predicates and learners within each size, with independent streams across sizes. Six new-C OR gains formed the primary family. Separate families covered six new-C XOR gains, four original-C gains, sixteen Ridge gains, twelve new-minus-original relative-gain differences and twelve corresponding absolute-reduction differences. Intervals condition on the fitted models, the realized calibration schedules and finite continuation banks.

Supplementary Section~\ref{supp:calibrationrobustness} reports the results and accompanying Source Data.

\vspace{-0.8\baselineskip}
\section*{Data availability}
The data supporting the findings of this study, including the source data for the figures and tables, are available in the \href{https://1drv.ms/f/c/07c8070870ea6501/IgDvqfqvxCB_SrRF76dXqlbhAcIF_2HU62YDgPeFrMtkxlo?e=MFvvMa}{QAOA data collection}.

\vspace{-0.8\baselineskip}
\section*{Author contributions}
S.K. conceived the project. M.K. developed the computational framework, carried out the simulations and analyses, and drafted the manuscript. S.K. supervised the research and provided theoretical guidance. All authors contributed to the interpretation of the results, reviewed and revised the manuscript, and approved the final version.

\vspace{-0.8\baselineskip}
\section*{ACKNOWLEDGMENTS}
We thank Dr. Travis Humble for many useful discussions. This work was supported by the U.S. Department of Energy (DOE), Office of Science, through the Quantum Science Center (QSC), a National Quantum Information Science Research Center.

\bibliography{references}
\onecolumngrid
\clearpage
\raggedbottom
\makeatletter
\setlength{\@fptop}{0pt}
\setlength{\@fpsep}{12pt}
\setlength{\@fpbot}{0pt plus 1fil}
\makeatother
\setcounter{section}{0}
\renewcommand{\thesection}{S\arabic{section}}
\setcounter{figure}{0}
\setcounter{table}{0}
\setcounter{equation}{0}
\renewcommand{\thefigure}{S\arabic{figure}}
\renewcommand{\thetable}{S\arabic{table}}
\renewcommand{\theequation}{S\arabic{equation}}
\renewcommand{\theHsection}{S\arabic{section}}
\renewcommand{\theHfigure}{S\arabic{figure}}
\renewcommand{\theHtable}{S\arabic{table}}
\renewcommand{\theHequation}{S\arabic{equation}}
\section{Supplementary information}
\label{supp:main}

\subsection{Evidence cohorts and their roles}
\label{supp:cohorts}

Table~\ref{supp:studyguide} distinguishes the main studies, their evaluation roles and analyses that reuse earlier cohorts. Study-specific inference families are defined in Methods; the physical contrasts and Pauli-weight extension are reported in Sections~\ref{supp:primary} and~\ref{supp:xorweights}.

\begin{table}[htbp]
\centering
\footnotesize
\setlength{\tabcolsep}{3.5pt}
\caption{Study designs and physical questions. All studies listed here cover $n=10,12,14,16$ except sample robustness at $n=10$ and calibration-budget robustness at $n=14,16$. Counts are independent instances per size; repeated depths, generators and continuations remain grouped by instance. Statistical families are specified in Methods.}
\label{supp:studyguide}
\begin{tabular}{llll}
\toprule
\manuscriptcell{0.17}{Study} & \manuscriptcell{0.18}{Instances per size} & \manuscriptcell{0.25}{Predictor design} & \manuscriptcell{0.30}{Question and evaluation role} \\
\midrule
\manuscriptcell{0.17}{Exact-component confirmation} & \manuscriptcell{0.18}{256 training; 512 new evaluation} & \manuscriptcell{0.25}{Separate scalar and component fits per recipe} & \manuscriptcell{0.30}{Which response components account for predictive gains? Independent confirmation.} \\[3pt]
\manuscriptcell{0.17}{Pauli-weight Max3XOR extension} & \manuscriptcell{0.18}{New 256 training; 512 evaluation} & \manuscriptcell{0.25}{Paired AC/CC; separate scalar and weight-sector fits} & \manuscriptcell{0.30}{Which Pauli weights carry complementary predictive information? New-cohort evaluation.} \\[3pt]
\manuscriptcell{0.17}{High-reward preparation} & \manuscriptcell{0.18}{New 256 training; 512 evaluation} & \manuscriptcell{0.25}{Paired A/H preparations and A/C continuations; common baseline} & \manuscriptcell{0.30}{Does information persist after preparing high-reward states? Independent paired comparison.} \\[3pt]
\manuscriptcell{0.17}{Seven-centre discovery} & \manuscriptcell{0.18}{256 training; 256 discovery evaluation} & \manuscriptcell{0.25}{Separate fits at each centre} & \manuscriptcell{0.30}{How does continuation reshape usefulness at a fixed state? Exploratory comparison.} \\[3pt]
\manuscriptcell{0.17}{18-condition normalized map} & \manuscriptcell{0.18}{Separate 256 training; 256 discovery evaluation} & \manuscriptcell{0.25}{Separate fits per condition} & \manuscriptcell{0.30}{Do cost and mixer shifts differ at matched displacement? Exploratory comparison.} \\[3pt]
\manuscriptcell{0.17}{Prospective transfer} & \manuscriptcell{0.18}{256 seven-centre training; 512 fresh evaluation} & \manuscriptcell{0.25}{Models trained on pooled data from seven centres, then frozen} & \manuscriptcell{0.30}{Does predictive information transfer to four unseen centres? Independent evaluation.} \\[3pt]
\manuscriptcell{0.17}{Z2 diagonal-summary control} & \manuscriptcell{0.18}{Exact-component cohort: 256 training; 512 evaluation} & \manuscriptcell{0.25}{New Z2 fits; B/M3 models unchanged} & \manuscriptcell{0.30}{Does diagonal aggregation explain M3's benefit? Retrospective control.} \\[3pt]
\manuscriptcell{0.17}{Component-predictor sensitivity} & \manuscriptcell{0.18}{512 exact-component evaluation} & \manuscriptcell{0.25}{ET scalar predictions fixed; Ridge component predictions substituted} & \manuscriptcell{0.30}{How does the fitted component predictor change the signed account? Retrospective sensitivity.} \\[3pt]
\manuscriptcell{0.17}{$n=10$ sample robustness} & \manuscriptcell{0.18}{Five new training cohorts of 256; 1,024 fresh evaluation} & \manuscriptcell{0.25}{Exact-component model and new cohort-specific fits} & \manuscriptcell{0.30}{Does XOR's benefit recur across training and evaluation samples? Independent evaluation.} \\[3pt]
\manuscriptcell{0.17}{Calibration-budget robustness ($n=14,16$)} & \manuscriptcell{0.18}{New 256 training; 512 evaluation; $3\times64$ calibration} & \manuscriptcell{0.25}{Separate B/M3 fits at original and three new C sets} & \manuscriptcell{0.30}{Does added information persist at matched calibration budgets? Fresh-cohort paired comparison.} \\
\bottomrule
\end{tabular}
\end{table}

Point estimates, interval endpoints, normalization conventions and original inference-family identifiers for the other tabulated diagnostics are supplied in Source Data, \nolinkurl{source_data/supplementary_diagnostics.tsv}. These entries retain their reported display precision; the associated full-precision result files remain available in the companion data collection.

Additional scalar-prediction studies summarized below---scalar confirmation, learning-size analysis and calibration transfer---complement the exact-component confirmation in Table~\ref{supp:studyguide}. Continuation diagnostics and prospective results appear in Supplementary Section~\ref{supp:transfer}.

The scalar confirmation generated 256 training and 512 evaluation instances at each of $n=10,12,14,16$, giving 3,072 instances. It evaluated OR AC and XOR CC under identity, local-$Z$, and selective-$ZZ$ transformations. Its primary Extra Trees comparisons used previously frozen models trained on 64 instances at $n=10,12$, and newly trained models with 256 instances at $n=14,16$. Accordingly, size-dependent changes in these primary results cannot be assigned solely to system size. The 55 primary contrasts assessed observable augmentation, OR--XOR differences, phase attenuation and the sufficiency of a single X summary. Fifty-two simultaneous intervals excluded zero in the specified direction; the other three concerned attenuation. The 24 OR improvements ranged from 16.96\% to 34.36\%. A separate secondary family contained 2,162 contrasts. This study used 5,000 paired bootstrap replicates, rather than the 4,000 used for exact components.

The learning-size study used five training cohorts and a common evaluation set of 2,048 instances per size. Increasing training size from 256 to 512 and 1,024 retained M3 benefits. At 1,024 training instances, the reported relative improvements were 23.64\% and 24.95\% for OR AC at $n=14,16$, and 10.14\% and 13.43\% for XOR CC. The common evaluation data, reused from earlier work in the study, support a matched comparison across training sizes.

A separate transfer study fixed the trained predictors before evaluating five newly generated calibration centres and new instances at $n=10,12$. All 40 primary contrasts supported the specified benefits and relative 5\% non-inferiority margins for the single X and XX summaries against M3. A distinct family of 40 contrasts supported AC--AA and AC--CC differences. This establishes transfer within the tested calibration-generation procedure; it does not extend the exact-component results to every calibration centre. Simultaneous intervals from separate studies are not combined into one study-wide 95\% guarantee.

\subsection{Sensitivity to the continuation sample bank}
\label{supp:bank}
We retained the saved model predictions and recalculated targets using all 64 continuations, the first 32, and the last 32. The final half was recovered as $y_{\mathrm{last}}=2y_{64}-y_{\mathrm{first}}$. Both halves retained every specified direction: all 14 intervals at $n=10$, all 22 jointly at $n=12,14$, and all 14 at $n=16$ excluded zero within their separate half-bank families. Each half contributes seven contrasts at $n=10$ and $n=16$, and eleven jointly at $n=12,14$. These nested halves share the original evaluation instances and fixed models.

Separate families of 21 paired differences at each of $n=10$ and $n=16$, and 33 jointly at $n=12,14$, compared each half with the full bank and the halves with each other. At $n=10$, only the two half-minus-full differences for normalized OR AC $T_m$ excluded zero, both downwards. At $n=12,14$, five last-half-minus-full contrasts had intervals below zero. At $n=16$, two first-half-minus-full contrasts had intervals below zero: normalized OR AC $T_m$ and its local-$Z$ attenuation. All between-half intervals included zero within each separate family, without establishing equivalence.

The normalized magnitudes changed partly because the baseline MSE was recomputed for each target. At $n=10$, $100T_m/R_B$ in OR AC changed from 45.71 to 42.28 and 42.87 in the first and last halves. Its absolute numerator changed by $-0.35\%$ and $+0.35\%$, while baseline MSE increased by 7.72\% and 6.98\%. For the five contrasts with a confirmed last-half decrease at $n=12,14$, absolute numerator changes ranged from approximately $-2.03\%$ to $+0.22\%$, whereas baseline-MSE increases ranged from 7.71\% to 23.31\%. At $n=16$, normalized OR AC $T_m$ changed from 28.95 to 26.51 and 27.45; its absolute numerator changed by $-0.22\%$ and $+0.22\%$, while baseline MSE increased by 8.95\% and 5.68\%. For any fixed predictor $f$, the identity
\begin{equation}
 \frac{\mathbb E[(y_{\mathrm{first}}-f)^2]+\mathbb E[(y_{\mathrm{last}}-f)^2]}{2}
 -\mathbb E[(y_{64}-f)^2]
 =\mathbb E[(y_{\mathrm{first}}-y_{64})^2]
 \label{supp:bankidentity}
\end{equation}
explains why half-bank target variation can increase the average denominator. Absolute residual-alignment terms are linear in the component targets and average back to their full-bank values. The quantitative primary analysis therefore retains the prespecified 64-continuation bank. This nested analysis quantifies fixed-model sensitivity to continuation sampling within the observed bank.

\begin{figure}[htbp]
\centering
\manuscriptfigure[0.74\textheight]{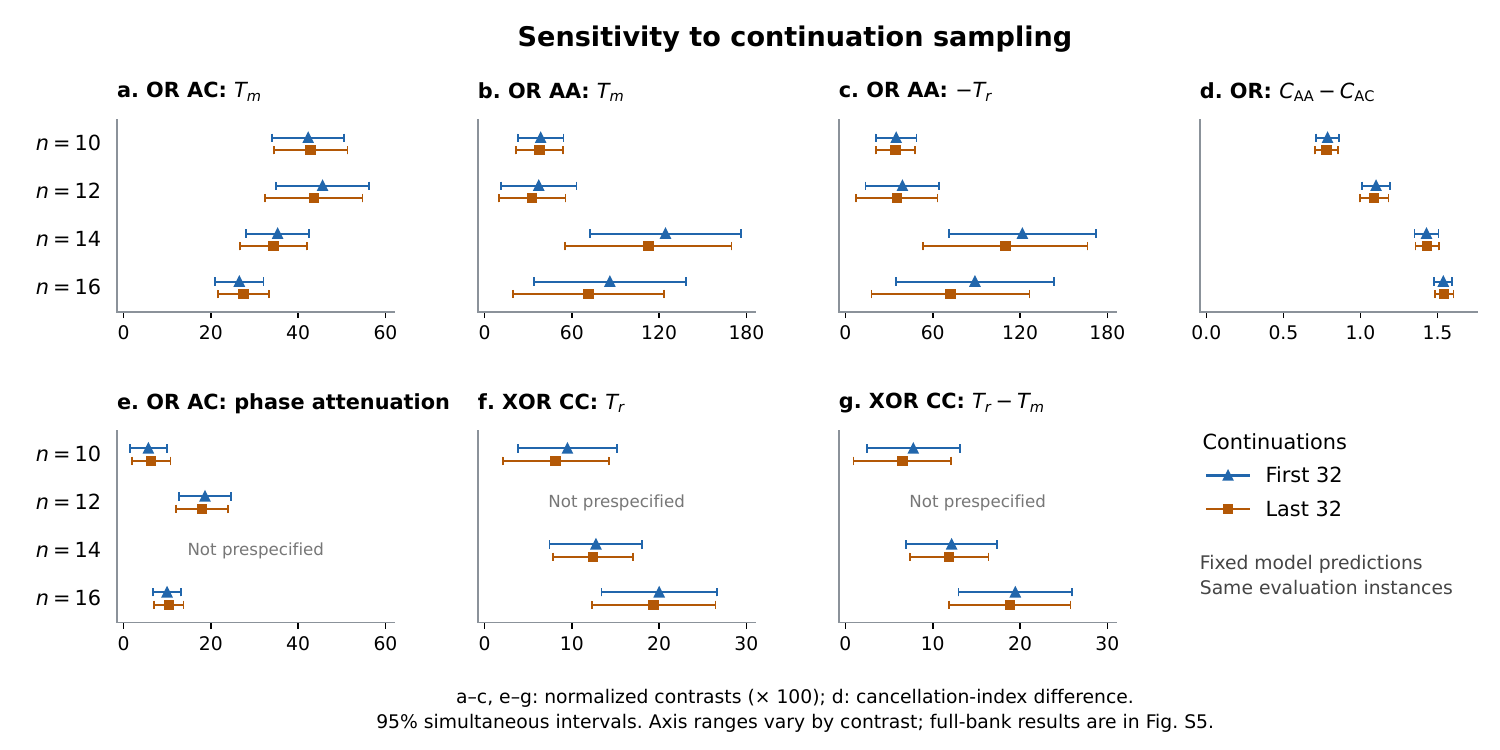}
\caption{\textbf{Fixed-model sensitivity to continuation sampling.} The first and last 32 continuations of the original bank are compared at $n=10,12,14,16$, with model predictions fixed. (a--c) Direct and opposing residual alignment in Max3SAT. (d) AA--AC cancellation-index difference. (e) Attenuation of direct alignment after local-$Z$ rotation. (f,g) Residual alignment and residual-minus-direct alignment in Max3XOR. Panels (a--c,e--g) show 21 contrasts per half, normalized by the corresponding half-bank baseline MSE and multiplied by 100; the phase contrast in (e) uses the identity baseline MSE as a common denominator. Axis ranges vary by contrast. ``Not prespecified'' marks comparisons outside that size's original contrast family. All 50 intervals retain the specified direction. The 95\% simultaneous intervals preserve separate 14-contrast ($n=10$), 22-contrast ($n=12,14$ jointly) and 14-contrast ($n=16$) families, with paired instance-level resampling. Full-bank results appear in Fig.~\ref{supp:confirmation}.}
\label{supp:bankfigure}
\end{figure}

\subsection{Sensitivity to the component predictor}
\label{supp:componentsensitivity}

The scalar response predictors determine the total prediction gain, whereas the component predictors allocate its signed terms in Eq.~\eqref{eq:loss_identity}. The retrospective Ridge-component comparison at $n=10,12,14,16$ holds the Extra Trees B and M3 scalar predictions fixed (Methods~\ref{meth:componentsensitivity}). Physical targets, scalar risks and gains, and the four physical cancellation contrasts therefore remain unchanged.

All 21 point estimates retain the originally specified directions; 18 have positive bounds in the single 21-contrast Ridge simultaneous family. The three intervals containing zero are at $n=10$: the local-$Z$ attenuation of OR AC direct alignment, the XOR complementary alignment, and the XOR complementary-minus-direct alignment. The separate 21-contrast Ridge$-$ET difference family detects changes in six contrasts. The main directional relationships therefore persist under this change of component predictor, with quantitative redistribution among the signed terms.

The $n=10$ XOR case illustrates why component prediction quality should accompany that redistribution. Replacing Extra Trees components by Ridge reduces $T_r/R_B$ from $0.09458$ to $0.01897$, while the correction term increases from $0.03505$ to $0.12218$. The complementary-component MSE divided by its target variance rises from $0.492$ to $0.849$. The unchanged scalar predictions have the same total improvement in both accounts, and all five terms recover that improvement.

\subsection{Mean offsets and fluctuation cancellation}
\label{supp:centering}
We performed a post hoc descriptive analysis of the saved exact-component targets to distinguish cancellation of mean offsets from cancellation of fluctuations. Each of $n=10,12,14,16$ contains 512 evaluation instances, with four paired depth--generator conditions per instance. All expectations give equal weight to instances and to these four conditions. The analysis uses the original 64-continuation targets and does not add contrasts to the prespecified inference family.

Writing $v=u+r$, $\mu_m=\mathbb E[m]$, $\mu_v=\mathbb E[v]$ and $D=\mathbb E[m^2]+\mathbb E[v^2]$, we separate the original index into
\begin{equation}
 C=C_{\mu}+C_{\mathrm{cov}},\qquad
 C_{\mu}=\frac{-2\mu_m\mu_v}{D},\qquad
 C_{\mathrm{cov}}=\frac{-2\operatorname{Cov}(m,v)}{D}.
 \label{supp:mean_cov_split}
\end{equation}
These two terms use the same denominator and sum to the original $C$. A distinct centred index,
\begin{equation}
 C_{\mathrm{ctr}}=\frac{-2\operatorname{Cov}(m,v)}
 {\operatorname{Var}(m)+\operatorname{Var}(v)},
 \label{supp:centered_index}
\end{equation}
normalizes only by fluctuation power. Its denominator differs from that of $C_{\mathrm{cov}}$.

Pooling depths and generators can itself induce covariance. Let $S$ denote the four fixed depth--generator conditions and $I$ the instance. We therefore also remove the mean separately within each $S$ and calculate
\begin{equation}
 C_{\mathrm{within}}=
 \frac{-2\mathbb E_S[\operatorname{Cov}_I(m,v\mid S)]}
 {\mathbb E_S[\operatorname{Var}_I(m\mid S)+\operatorname{Var}_I(v\mid S)]}.
 \label{supp:within_index}
\end{equation}
This is a ratio of averaged moments, rather than an average of the four index values. Empirical covariances use the expectation convention with divisor $N$. The decomposition
$\operatorname{Cov}(m,v)=\mathbb E_S[\operatorname{Cov}_I(m,v\mid S)]+\operatorname{Cov}_S(\mathbb E_I[m\mid S],\mathbb E_I[v\mid S])$
separates variation within conditions from differences between condition means.

Using the same original denominator $D$ for all three contributions gives
\begin{equation}
\begin{gathered}
 C=C_{\mu}+C_{\mathrm{between}}^{(D)}+C_{\mathrm{within}}^{(D)},\\
 C_{\mathrm{between}}^{(D)}=
 \frac{-2\operatorname{Cov}_S(\mathbb E_I[m\mid S],\mathbb E_I[v\mid S])}{D},\qquad
 C_{\mathrm{within}}^{(D)}=
 \frac{-2\mathbb E_S[\operatorname{Cov}_I(m,v\mid S)]}{D}.
\end{gathered}
\label{supp:threeway_cancellation}
\end{equation}
The superscript $(D)$ distinguishes the additive within-condition contribution from the fluctuation-normalized index $C_{\mathrm{within}}$ in Eq.~\eqref{supp:within_index}.

\begin{table}[htbp]
\centering
\small
\setlength{\tabcolsep}{9pt}
\caption{Physical cancellation before and after centering in the matched OR conditions. Values are descriptive estimates from the saved 64-continuation targets. The mean contribution $C_\mu$ uses the original denominator $D$; $C_{\mathrm{ctr}}$ and $C_{\mathrm{within}}$ use pooled and within-condition fluctuation variances, respectively. Full moment decompositions and the local-$Z$/XOR diagnostics are supplied in Source Data.}
\label{supp:centeringtable}
\begin{tabular}{lrrrrr}
\toprule
Condition & $n$ & $C$ & $C_\mu$ & $C_{\mathrm{ctr}}$ & $C_{\mathrm{within}}$\\
\midrule
OR AA & 10 & 0.864 & -0.009 & 0.910 & 0.916\\
 & 12 & 0.932 & 0.116 & 0.943 & 0.946\\
 & 14 & 0.980 & 0.217 & 0.981 & 0.982\\
 & 16 & 0.986 & 0.277 & 0.983 & 0.983\\
\midrule
OR AC & 10 & 0.083 & 0.087 & -0.011 & -0.061\\
 & 12 & -0.162 & -0.091 & -0.187 & -0.206\\
 & 14 & -0.453 & -0.300 & -0.337 & -0.308\\
 & 16 & -0.559 & -0.385 & -0.315 & -0.263\\
\bottomrule
\end{tabular}
\end{table}

OR AA retains strong cancellation after both forms of centering at $n=10,12,14,16$ (Table~\ref{supp:centeringtable}). The same-denominator fraction $100C_{\mathrm{within}}^{(D)}/C$ is 99.82\%, 87.08\%, 75.79\% and 69.35\% at $n=10,12,14,16$, respectively (Eq.~\eqref{supp:threeway_cancellation}; Source Data). This contribution fraction uses the original $D$ and is distinct from the within-condition fluctuation index in Table~\ref{supp:centeringtable}. In contrast, the positive uncentred index of OR AC at $n=10$ is dominated by its mean-offset term, while its pooled and within-condition centred indices are negative. The same sign change occurs for OR AC with local Z at $n=12$. Mean offsets and fluctuation cancellation thus distinguish different aspects of the response ensemble.

The more detailed $n=12,14$ checks resolve individual conditions and half-bank centering. Across the four depth--generator conditions, the centred AA index ranges from 0.894 to 0.980 at $n=12$ and from 0.965 to 0.991 at $n=14$; the corresponding OR AC indices are negative in every condition. Across the two 32-continuation halves, aggregate AA $C_{\mathrm{within}}$ is 0.945--0.947 at $n=12$ and 0.980--0.982 at $n=14$.

\subsection{Interpretation and numerical checks}
\label{supp:interpretation}
Relative gains, cancellation indices and sector ratios are defined only for positive denominators; zero-denominator cases are undefined. All reported estimates satisfy this positivity condition.

The physical cancellation index is based on uncentred second moments. Since $c_2=m+u+r$,
\begin{equation}
 C=1-\frac{\mathbb E[c_2^2]}{\mathbb E[m^2]+\mathbb E[(u+r)^2]}.
 \label{supp:cidentity}
\end{equation}
A large value describes cancellation in the specified physical decomposition. The empirical contrast is that the same feature-linked Pauli decomposition gives substantially different cancellation in AA and AC. Sensitivity of the signed loss account to the component predictor is assessed in Section~\ref{supp:componentsensitivity}.

We checked the exact response decomposition and five-term loss identity on the stored targets and predictions. Independent small-matrix tests at $n=2$--5 compared good-set-projected state-vector propagation with an explicitly formed double commutator, streamed Pauli coefficients with direct trace evaluations, and the complementary component with a direct sum of Pauli terms outside the selected subspace. These tests included empty and full good sets and cost-angle reflection. CPU and accelerator implementations were also compared for full-rank coefficient calculations and prospective response generation. Agreement was at absolute errors of order $10^{-15}$ or smaller, validating both the algebraic identities and their numerical implementation.

\subsection{Cost-angle reflection and matched-state controls}
\label{supp:symmetry}
Write the continuation as a product of real-diagonal cost evolutions and X mixers, and let $\mathcal Z=\prod_qZ_q$. Since $\mathcal Z C\mathcal Z=C$ and $\mathcal Z(\sum_qX_q)\mathcal Z=-\sum_qX_q$, entrywise conjugation gives
\begin{equation}
 V(-\boldsymbol\gamma,\boldsymbol\beta)
 =\mathcal Z V(\boldsymbol\gamma,\boldsymbol\beta)^*\mathcal Z.
 \label{supp:Vreflection}
\end{equation}
Entrywise conjugation preserves the gate order. Both the diagonal intervention generator and good-solution projector commute with $\mathcal Z$ and are real, which yields Eq.~\eqref{eq:gamma_reflection}. For a Pauli string $Q$ with $N_X$ X factors and $N_Y$ Y factors,
$\mathcal ZQ\mathcal Z=(-1)^{N_X+N_Y}Q$ and $Q^*=(-1)^{N_Y}Q$. The real coefficient $\alpha_Q=2^{-n}\operatorname{Tr}(KQ)$ therefore satisfies
\begin{equation}
 \alpha_Q(-\boldsymbol\gamma,\boldsymbol\beta)
 =(-1)^{N_X}\alpha_Q(\boldsymbol\gamma,\boldsymbol\beta).
 \label{supp:alphareflection}
\end{equation}
For the selected blocks, X changes sign and XX/YY do not. At a fixed prepared state, the block-average contribution consequently obeys
\begin{equation}
 m(\boldsymbol\gamma,\boldsymbol\beta)-m(-\boldsymbol\gamma,\boldsymbol\beta)
 =2m_X(\boldsymbol\gamma,\boldsymbol\beta).
 \label{supp:mreflection}
\end{equation}
When every actual cost angle is zero, every X coefficient vanishes; XX and YY need not vanish. The interpolated midpoint in the discovery study includes nonzero sampled deviations and is not the exact zero-cost-angle control.

Numerical symmetry checks used reserved $n=12,14$ pilot instances, both depths and both generators. Reflection changed every actual cost angle, including its deviation, while keeping mixer angles and the prepared state fixed; the zero-angle control removed both cost-angle centres and deviations. Full-rank calculations in double precision verified the coefficient identities and zero-X condition with the numerical accuracy summarized in Supplementary Section~\ref{supp:interpretation}.

\subsection{Continuation discovery and transfer diagnostics}
\label{supp:transfer}
This section compares physical continuation diagnostics, secondary transfer contrasts and fixed-prediction bank sensitivity at $n=10,12,14,16$ for the designs in Methods~\ref{meth:transfer}.

\begin{figure}[htbp]
 \centering
 \hspace*{-10pt}\scalebox{0.90}{\manuscriptfigure[0.65\textheight]{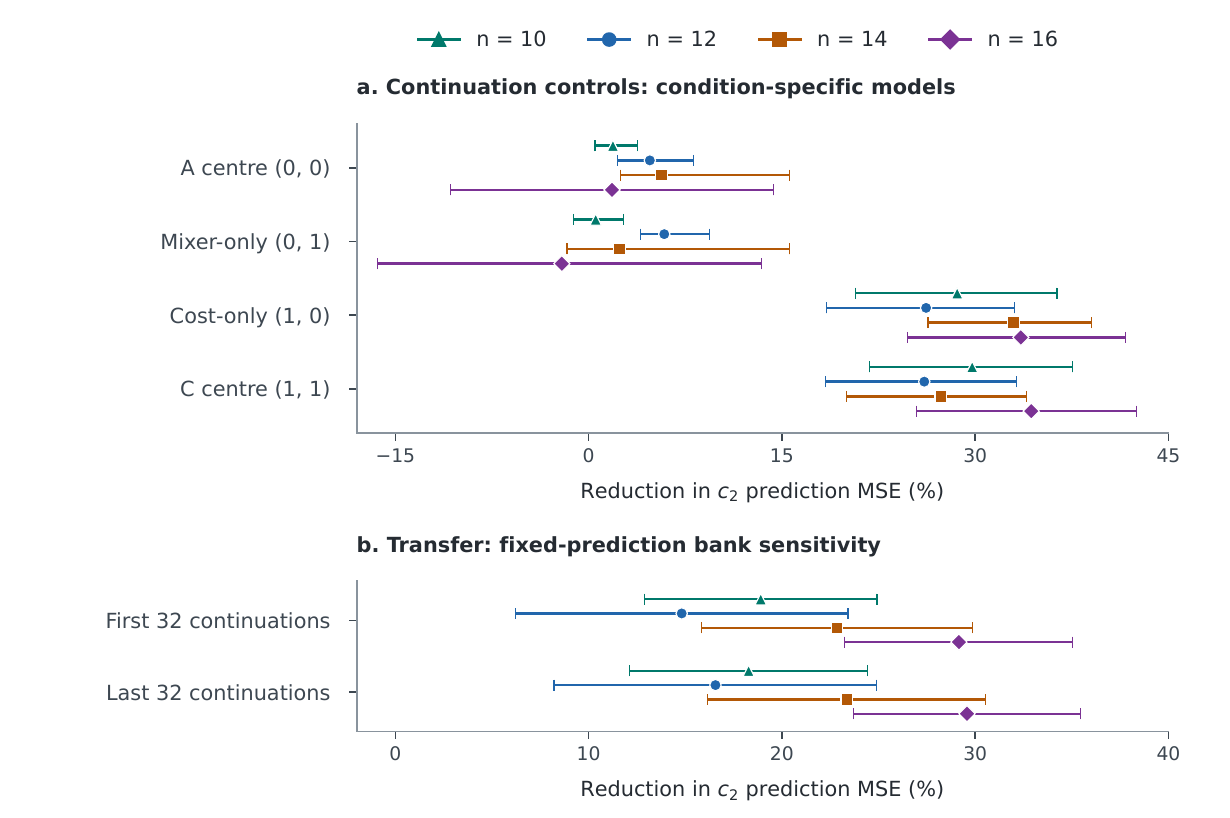}}\hspace*{10pt}
 \caption{\textbf{Continuation controls and fixed-prediction sensitivity.}
 (a) Extra Trees/M3 gains at the A and C endpoints, and for mixer-only and cost-only changes, at $n=10,12,14,16$. These four controls complement the diagonal interpolation in Fig.~\ref{fig:continuationphysics}a. The coordinates are the cost- and mixer-angle interpolation fractions $(a,b)$; all points share the same prepared state within each instance and depth, obtained by A-initialized preparation with 40 optimization updates. Each condition has a separately fitted predictor and 256 discovery-evaluation instances per size. Intervals are exploratory, unadjusted 95\% percentiles from 1,000 paired instance-bootstrap replicates. (b) Gains in the prospective transfer evaluation when the original fixed predictions are compared with targets formed from the first or second 32 continuations. The 95\% simultaneous intervals retain separate secondary families of 23 comparisons each at $n=10$ and $n=16$, and 46 jointly at $n=12,14$, using 4,000 paired instance-bootstrap replicates. The halves share their evaluation instances and are sensitivity analyses, not additional independent cohorts.}
 \label{supp:continuationfigure}
\end{figure}

The saved physical diagnostics for all four sizes and all seven discovery conditions, shown across Figs.~\ref{fig:continuationphysics} and~\ref{supp:continuationfigure}a, are supplied in Source Data, \nolinkurl{source_data/continuation_physical_diagnostics.tsv}. For each condition, let $\alpha_{X_i,t}$ denote the coefficient of $X_i$ for continuation draw $t$, with the instance $I$ and depth--generator condition $S$ implicit. We report the signed block mean and the mean per-draw block RMS,
\begin{equation}
 \bar\alpha_X=\mathbb E_{I,S}\!\left[\frac{1}{64}\sum_{t=1}^{64}
 \frac{1}{n}\sum_{i=1}^n\alpha_{X_i,t}\right],\qquad
 \rho_X=\mathbb E_{I,S}\!\left[\frac{1}{64}\sum_{t=1}^{64}
 \left(\frac{1}{n}\sum_{i=1}^n\alpha_{X_i,t}^{\,2}\right)^{1/2}\right].
\label{supp:coefficient_diagnostics}
\end{equation}
The expectation weights the 256 discovery-evaluation instances and four depth--generator conditions equally. For $\rho_X$, the RMS across the $n$ coefficients is taken for each continuation draw before averaging the 64 draws and then the evaluation rows. The response components entering $C_{\mathrm{within}}$ are first averaged over the 64 continuations; the index then uses the within-condition moments in Eq.~\eqref{supp:within_index}.

At the midpoint, $\rho_X$ is 30.92\%, 30.84\%, 30.95\% and 32.12\% of the mean endpoint value at $n=10,12,14,16$, respectively. Its reduced magnitude accompanies the sign change in $\bar\alpha_X$. The midpoint also has $C_{\mathrm{within}}$ close to zero and near-zero M3 gain, showing why cancellation alone does not order the summaries' incremental predictive value. These are descriptive diagnostics of the discovery sweep.

Ridge/M3 retains positive overall gains of 5.00--5.92\%, and both continuation-bank halves retain positive Extra Trees gains (Fig.~\ref{supp:continuationfigure}b). These secondary comparisons retain separate simultaneous families of 23 contrasts each at $n=10$ and $n=16$, and 46 jointly at $n=12,14$; their estimates and interval endpoints are supplied in Source Data. The high-minus-low cost-angle contrast uses $a=0.875$ minus $a=0.125$, pooling raw loss across mixer-angle fractions before forming each gain. At $n=12,14$, all four point estimates are positive and the simultaneous intervals resolve two of them; at both $n=10$ and $n=16$, the Ridge and Extra Trees contrasts have positive bounds. Here high and low refer to the interpolation fraction $a$, so the contrast compares C-nearer with A-nearer centres rather than ordering the magnitudes of the actual cost angles.

Table~\ref{supp:transferlosses} gives the pooled absolute risks and their concentration across angle conditions. Across the four sizes, the corresponding baseline and M3 $R^2$ scores range from 0.7296--0.7752 and 0.8045--0.8191, respectively, with each score defined by $1-R/\operatorname{Var}(c_2)$. Complete conditionwise losses, gains and size-specific $R^2$ values are supplied in Source Data.

With equally weighted observations in all four angle conditions, the two $a=0.875$ conditions account for 98.45--99.93\% of the pooled baseline squared loss (Table~\ref{supp:transferlosses}). This is a share of prediction loss, not a share of instances: the prespecified gain pools raw squared losses before taking their ratio, rather than averaging conditionwise gain percentages. It is consequently close to the high-$a$ gains, while Fig.~\ref{fig:transfer} resolves predictive usefulness at each centre separately. At $n=16$, the A-nearer cost centre with the C-nearer mixer centre, $(a,b)=(0.125,0.875)$, gives a 24.05\% gain [14.96,33.14]\% despite its small absolute loss scale. Thus the conditionwise results identify useful information in regimes contributing little to the pooled loss.

\begin{table}[htbp]
\centering
\small
\setlength{\tabcolsep}{10pt}
\caption{Absolute scale and angle-condition weighting of the prospective transfer result. The risks pool the four unseen centres, depths and generators over 512 fresh instances per size. Both risk columns are in units of $10^{-4}$. The last column is the percentage of baseline squared loss contributed by the two C-nearer cost-angle centres; each centre has equal observation weight. Entries are descriptive estimates.}
\label{supp:transferlosses}
\begin{tabular}{rrrr}
\toprule
$n$ & $R_B\ (10^{-4})$ & $R_{\mathrm{M3}}\ (10^{-4})$ & Loss share at $a=0.875$ (\%)\\
\midrule
10 & 3.120031 & 2.510648 & 98.45\\
12 & 1.792662 & 1.495716 & 99.59\\
14 & 1.035200 & 0.782467 & 99.79\\
16 & 0.713501 & 0.491532 & 99.93\\
\bottomrule
\end{tabular}
\end{table}

\subsection{Generator-normalized continuation map}
\label{supp:matched}
\begin{figure}[htbp]
 \centering
 \manuscriptfigure[0.56\textheight]{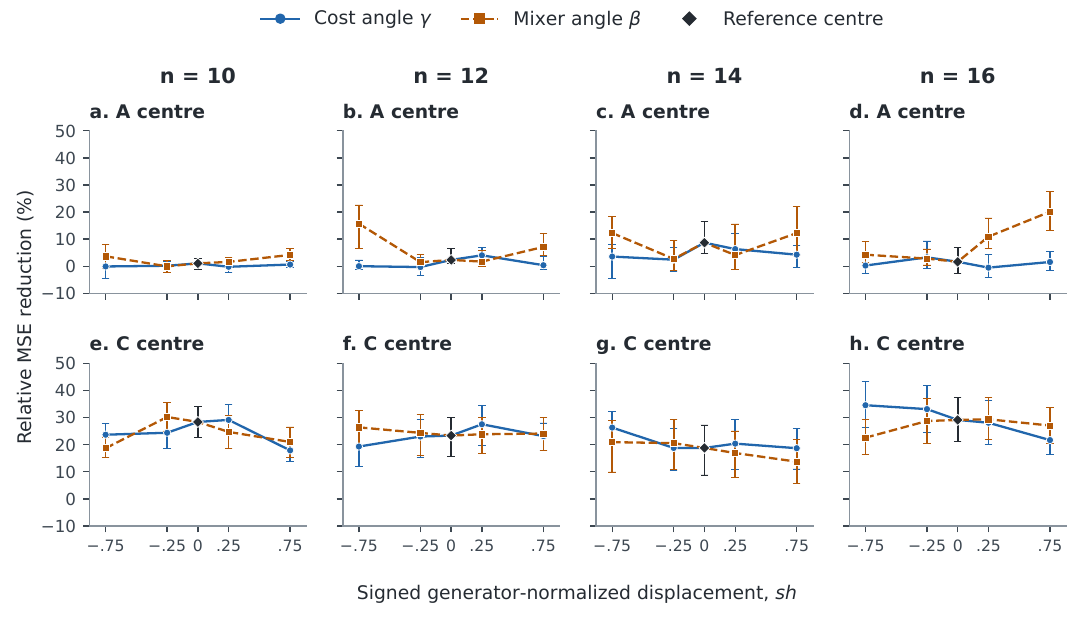}
 \caption{\textbf{Predictive gains under generator-normalized continuation changes.}
 Extra Trees/M3 gains on 256 discovery-evaluation instances per size, using condition-specific models. Columns distinguish $n=10,12,14,16$ and rows distinguish A and C continuation centres. The horizontal coordinate is signed displacement $s h$ in Eq.~\eqref{meth:matched_shift}. Blue circles show cost-angle changes; brown squares show mixer-angle changes. The black diamond at zero is the shared baseline centre, plotted once. Connecting lines guide the eye between the sampled conditions. Intervals are unadjusted 95\% percentiles from 1,000 paired instance-bootstrap replicates; all repeats and conditions share resampling weights within each size. Prepared states, generators and 64 continuation deviations are paired across conditions.}
 \label{supp:matchedfigure}
\end{figure}

The separate 18-condition map varies cost or mixer angles around A and C with the matched generator-weighted displacement in Eq.~\eqref{meth:matched_shift}. Figure~\ref{supp:matchedfigure} reports every Extra Trees/M3 condition at all four sizes. The common prepared state isolates the change of continuation within each instance; the normalization supplies a shared scale for the two angle sectors.

Source Data, \texttt{normalized\_continuation\_all\_sizes.tsv}, supplies all 18 conditions per size for Ridge/M3 and Extra Trees/M3, including relative and absolute MSE reductions, prediction risks and the exploratory intervals defined in Fig.~\ref{supp:matchedfigure}.

Table~\ref{supp:matchedcancellation} compares A with the two mixer shifts at $h=0.75$. At all four sizes, the negative shift lowers $C_{\mathrm{within}}$ and the positive shift raises it, while both increase the relative gain and absolute reduction $\Delta R=R_B-R_{\mathrm{M3}}$. The augmented predictor's MSE rises for the negative shift at all four sizes and falls for the positive shift at $n=12,14,16$; at $n=10$ it rises for both shifts. Thus increased incremental benefit can accompany either lower or higher prediction risks. These are descriptive contrasts; the conditionwise intervals in Fig.~\ref{supp:matchedfigure} do not provide intervals for their differences.

\begin{table}[htbp]
\centering
\small
\caption{Cancellation and the benefit of M3 augmentation under opposite mixer shifts around A. Values are descriptive estimates for Extra Trees/M3. The two risks and their difference $\Delta R$ are in units of $10^{-8}$; gain is $100\Delta R/R_B$. Values are rounded separately.}
\label{supp:matchedcancellation}
\begin{tabular}{@{}c@{\hspace{0.025\linewidth}\hspace{0.3pt}\hspace{0.025\linewidth}}c@{}}
\begin{minipage}[t]{0.47\linewidth}
\vspace{0pt}
\centering\small\setlength{\tabcolsep}{4pt}
\begin{tabular*}{\linewidth}{@{\extracolsep{\fill}}rrrrrr@{}}
\toprule
\multicolumn{6}{c}{$n=10$}\\
$s h$ & $R_B$ & $R_{\mathrm{M3}}$ & $\Delta R$ & Gain (\%) & $C_{\mathrm{within}}$\\
\midrule
$0$ & 383.250 & 379.109 & 4.141 & 1.08 & 0.9304\\
$-0.75$ & 914.544 & 880.881 & 33.663 & 3.68 & 0.5579\\
$+0.75$ & 750.289 & 718.779 & 31.510 & 4.20 & 0.9625\\
\bottomrule
\end{tabular*}
\end{minipage} & \begin{minipage}[t]{0.47\linewidth}
\vspace{0pt}
\centering\small\setlength{\tabcolsep}{4pt}
\begin{tabular*}{\linewidth}{@{\extracolsep{\fill}}rrrrrr@{}}
\toprule
\multicolumn{6}{c}{$n=12$}\\
$s h$ & $R_B$ & $R_{\mathrm{M3}}$ & $\Delta R$ & Gain (\%) & $C_{\mathrm{within}}$\\
\midrule
$0$ & 188.785 & 184.350 & 4.435 & 2.35 & 0.9266\\
$-0.75$ & 442.581 & 373.049 & 69.532 & 15.71 & 0.6166\\
$+0.75$ & 129.778 & 120.514 & 9.264 & 7.14 & 0.9794\\
\bottomrule
\end{tabular*}
\end{minipage}\\[8pt]
\begin{minipage}[t]{0.47\linewidth}
\vspace{0pt}
\centering\small\setlength{\tabcolsep}{4pt}
\begin{tabular*}{\linewidth}{@{\extracolsep{\fill}}rrrrrr@{}}
\toprule
\multicolumn{6}{c}{$n=14$}\\
$s h$ & $R_B$ & $R_{\mathrm{M3}}$ & $\Delta R$ & Gain (\%) & $C_{\mathrm{within}}$\\
\midrule
$0$ & 25.885 & 23.632 & 2.253 & 8.70 & 0.9686\\
$-0.75$ & 98.974 & 86.833 & 12.141 & 12.27 & 0.6782\\
$+0.75$ & 22.268 & 19.545 & 2.724 & 12.23 & 0.9905\\
\bottomrule
\end{tabular*}
\end{minipage} & \begin{minipage}[t]{0.47\linewidth}
\vspace{0pt}
\centering\small\setlength{\tabcolsep}{4pt}
\begin{tabular*}{\linewidth}{@{\extracolsep{\fill}}rrrrrr@{}}
\toprule
\multicolumn{6}{c}{$n=16$}\\
$s h$ & $R_B$ & $R_{\mathrm{M3}}$ & $\Delta R$ & Gain (\%) & $C_{\mathrm{within}}$\\
\midrule
$0$ & 6.289 & 6.185 & 0.104 & 1.65 & 0.9870\\
$-0.75$ & 60.318 & 57.739 & 2.579 & 4.28 & 0.5667\\
$+0.75$ & 2.996 & 2.393 & 0.603 & 20.12 & 0.9967\\
\bottomrule
\end{tabular*}
\end{minipage}
\end{tabular}
\end{table}

For a centre and radius $h$, let $c_{2,0}$ be its response and $c_{2,\xi,s}$ the response after a shift in sector $\xi\in\{\gamma,\beta\}$ with sign $s$. Each response is averaged over the paired bank of 64 continuations. Define
\begin{equation}
 D_\xi=\frac12\sum_{s=\pm1}\mathbb E\bigl[(c_{2,\xi,s}-c_{2,0})^2\bigr],
 \qquad
 G_\xi=\left[\frac12\sum_{s=\pm1}(g_{\xi,s}-g_0)^2\right]^{1/2},
 \label{supp:matched_metrics}
\end{equation}
where the expectation weights instances and their four depth--generator observations equally, and $g=1-R_{\mathrm{M3}}/R_B$ is the corresponding conditionwise gain. The sector ratios $D_\beta/D_\gamma$ and $G_\beta/G_\gamma$ are supplied in Source Data for all sizes, centres and radii. Mixer shifts give larger mean-square response changes in all sixteen settings. The larger gain-change RMS instead depends on size, centre and radius. $G_\xi$ measures the magnitude of a gain change, including a decrease, and its definition uses the baseline risk at each condition. Thus the sector that produces the larger physical response displacement need not produce the larger change in the summaries' incremental predictive value. These descriptive ratios carry no inferential intervals.

\subsection{Response distributions and individual predictions}
\label{supp:rawpredictions}
\begin{figure*}[!htbp]
 \centering
 \manuscriptfigure[0.80\textheight]{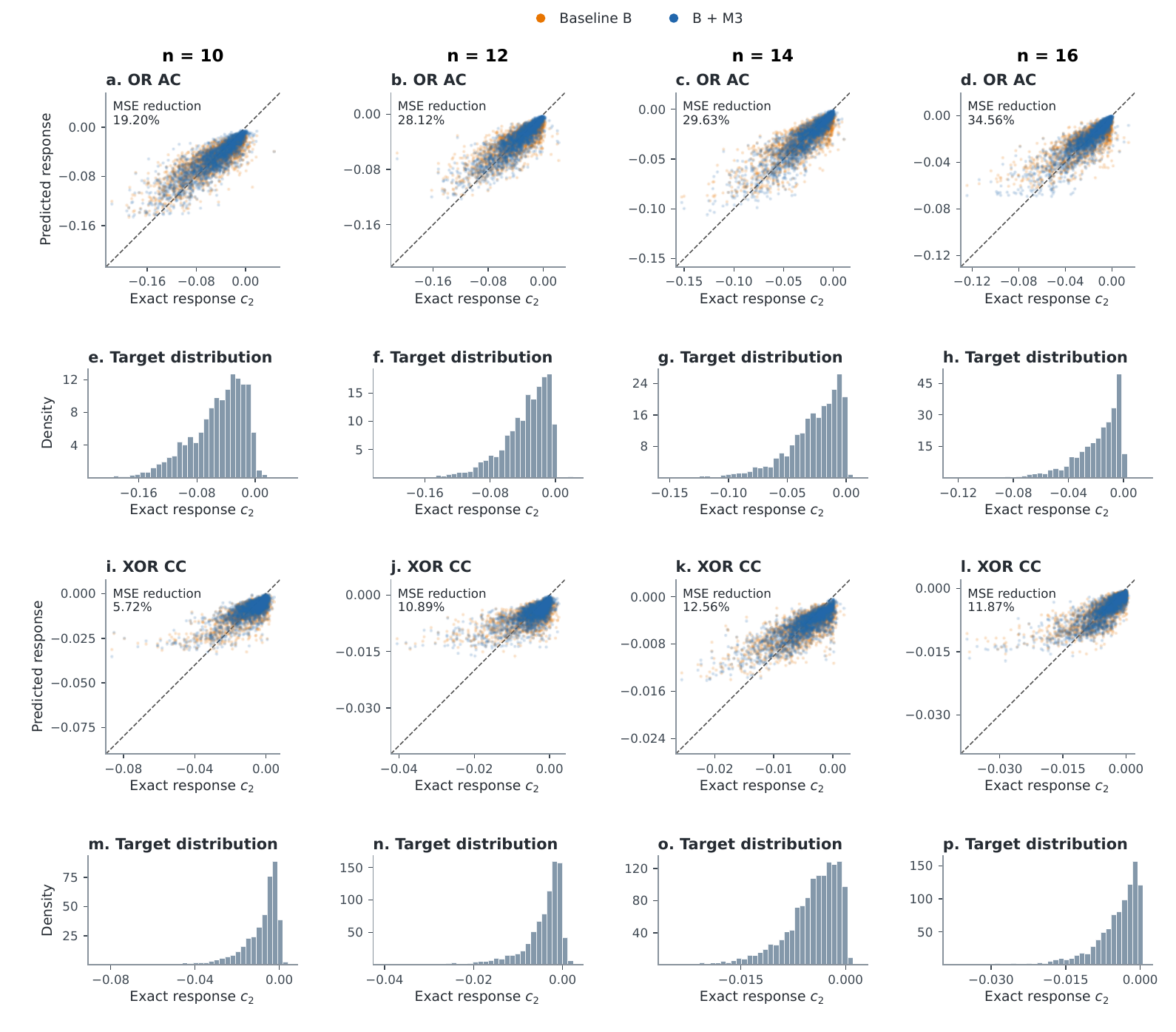}
 \caption{\textbf{Individual response predictions and target distributions.}
 Columns show $n=10,12,14,16$. The first two rows show OR AC and the last two show XOR CC, using each size's exact-component evaluation cohort.
 \textbf{a--d, i--l}, Exact response targets versus predictions from the fixed Extra Trees baseline $B$ (orange) and its M3 augmentation (blue). The dashed line marks equality; the annotation reports the MSE reduction from adding M3.
 \textbf{e--h, m--p}, Distributions of the same response targets. Each condition includes all 2,048 depth--generator observations from 512 independent evaluation instances, with four correlated observations per instance. Targets are exact-operator responses averaged over the fixed bank of 64 continuations. These plots reuse the saved predictions without further fitting or selection.}
 \label{supp:rawpredictionfigure}
\end{figure*}

Figure~\ref{supp:rawpredictionfigure} displays the individual predictions underlying the OR AC and XOR CC gains in the exact-component evaluation cohorts.

\subsection{Direct response from the complete selected Pauli subspace}
\label{supp:selectedspace}
The physical decomposition distinguishes the block-mean response $m$ from the complete response $m+u$ carried by the selected Pauli subspace. To check whether a small $m$ conceals a large within-block response, we compute
\begin{equation}
 Q_{\mathcal P}=\frac{\mathbb E[(m+u)^2]}{\mathbb E[c_2^2]}
 =\frac{\mathbb E[m^2]+\mathbb E[u^2]+2\mathbb E[mu]}{\mathbb E[c_2^2]}.
 \label{supp:selectedspace_ratio}
\end{equation}
For the exact-component XOR CC cohorts, $Q_{\mathcal P}$ is 0.8890\%, 0.2561\%, 0.0861\% and 0.0376\% at $n=10,12,14,16$, respectively. The small full selected-space response, including $u$, coexists with M3 prediction gains (Figs.~\ref{fig:accounts}f and~\ref{fig:gain}). $Q_{\mathcal P}$ is a second-moment ratio for the specified X and clause-supported XX/YY subspace, not an additive causal share or an explained-variance fraction. It uses the saved moments under each size's original calibration.

\subsection{Pauli-weight-resolved information in Max3XOR}
\label{supp:xorweights}
The paired Pauli-weight AC/CC cohorts in Methods~\ref{meth:xorweights} locate the complementary predictive information by Pauli weight (Fig.~\ref{fig:xorweights}). Table~\ref{supp:xorweighttable} reports the prespecified scalar gains and higher-weight residual alignments, retaining each study's eight-effect inference family.

\begin{table}[htbp]
\centering
\small
\setlength{\tabcolsep}{6pt}
\caption{Max3XOR scalar prediction gains and high-weight residual alignment for Extra Trees/M3. Entries give estimates and approximate 95\% simultaneous intervals, in per cent of scalar baseline MSE. Each horizontal block retains its own eight-effect family: four scalar gains and four alignment terms. Each size uses 512 evaluation instances with four repeated depth--generator rows per preparation.}
\label{supp:xorweighttable}
\begin{tabular}{clrr}
\toprule
$n$ & Condition & Scalar gain (\%) & $100T_{\mathrm{hi}}/R_B$\\
\midrule
10 & CC & 3.98 [-1.53, 9.48] & 7.93 [2.91, 12.96] \\
10 & AC & 14.39 [8.30, 20.48] & 15.10 [9.81, 20.40] \\
12 & CC & 7.66 [3.21, 12.10] & 11.69 [7.46, 15.91] \\
12 & AC & 8.43 [3.74, 13.13] & 9.20 [4.55, 13.86] \\
\midrule
14 & CC & 10.04 [6.68, 13.39] & 14.43 [10.61, 18.24] \\
14 & AC & 19.19 [12.96, 25.41] & 20.39 [13.95, 26.83] \\
16 & CC & 13.54 [8.62, 18.46] & 21.02 [15.26, 26.79] \\
16 & AC & 8.12 [4.37, 11.87] & 7.26 [3.83, 10.69] \\
\bottomrule
\end{tabular}
\end{table}

Residual alignment measures how the scalar prediction change tracks the baseline component residual. A separate diagnostic evaluates the MSE of $r_{\mathrm{hi}}$ using sums of the jointly fitted component outputs. This diagnostic gives positive pointwise 95\% bounds in seven conditions, with Extra Trees reductions of 6.31--15.67\%. At $n=10$ CC the reduction is $-2.16\%$ (pointwise interval $[-6.81,2.27]\%$), alongside the positive scalar residual alignment above. The scalar loss identity combines all sector alignments with the scalar--component correction and prediction-change penalty. The refined physical decomposition closes to within $1.27\times10^{-15}$ across the training and evaluation data.

Ridge scalar gains range from 2.13\% to 11.72\%, and both these gains and the high-weight alignments have positive pointwise 95\% lower bounds in all eight conditions. Pauli weight specifies operator support size; it does not by itself diagnose entanglement or irreducible many-body correlations. Full-precision estimates, inference-family identifiers, component risks, weight spectra and physical moment matrices are supplied with Fig.~\ref{fig:xorweights} in Source Data.

\subsection{Independent confirmation of the physical accounts}
\label{supp:physicalconfirmation}

\label{supp:primary}
\label{sec:n10-extension}
\begin{figure*}[tbp]
 \centering
 \hspace*{-10pt}\scalebox{0.90}{\manuscriptfigure[0.70\textheight]{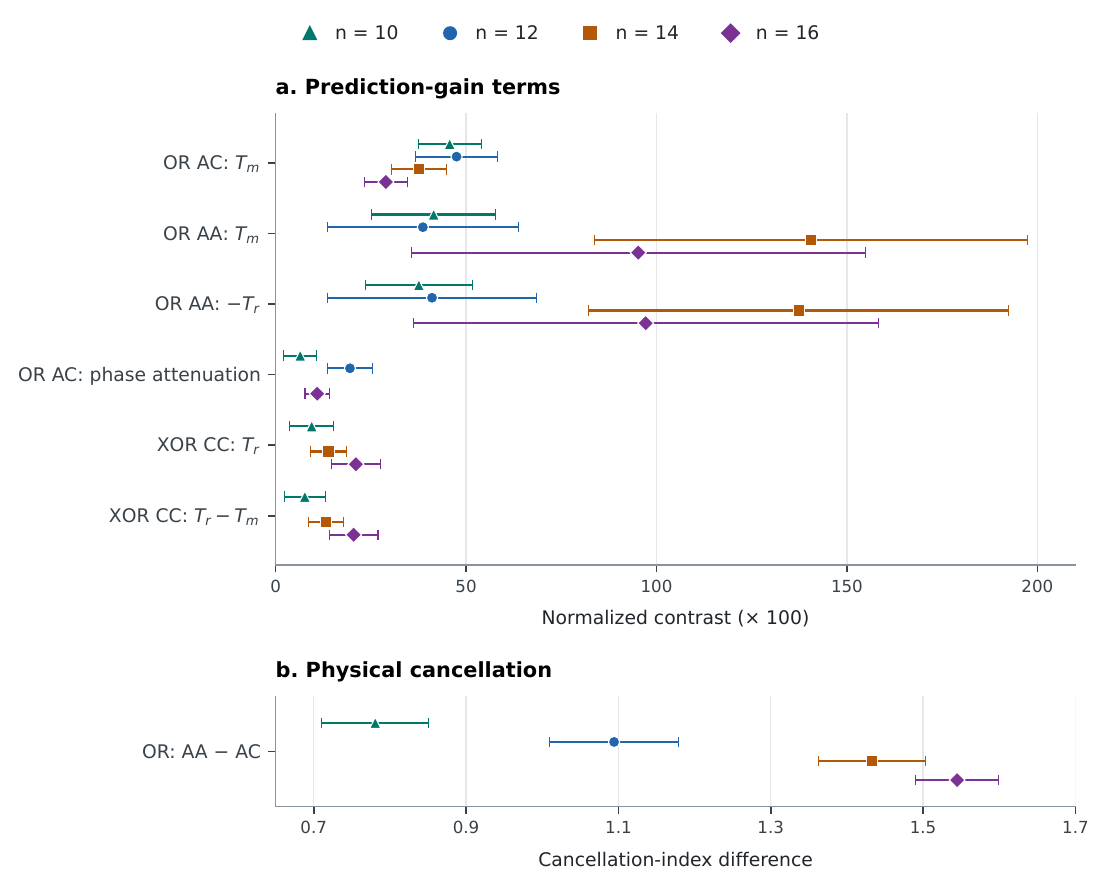}}\hspace*{10pt}
 \caption{\textbf{Independent confirmation of the prespecified physical accounts.}
 Twenty-five primary contrasts on 512 fresh evaluation instances per size retain three separate 95\% simultaneous families: seven contrasts at $n=10$, eleven jointly at $n=12,14$, and seven at $n=16$. Results use Extra Trees/M3 and 4,000 paired instance-level bootstrap resamples. (a) Twenty-one signed loss-term contrasts, grouped by contrast type and size, normalized by baseline MSE and multiplied by 100. OR AA complementary terms are sign reversed; the phase contrast uses the identity baseline MSE. Missing size markers indicate contrasts not prespecified in that primary family. (b) Four dimensionless AA--AC differences in physical cancellation. All intervals have the prespecified sign within their respective families. Each resampled instance carries its realized continuation draws and paired conditions; training data, fitted models and calibration centres are held fixed.}
 \label{supp:confirmation}
\end{figure*}

All 25 prespecified physical contrasts have 95\% simultaneous intervals in their predicted direction within their respective families on 2,048 new evaluation instances at $n=10,12,14,16$ (Fig.~\ref{supp:confirmation}). The contrasts, primary Extra Trees/M3 model and statistical procedure were fixed before each study's final instances were generated. The 11 contrasts assessed jointly at $n=12,14$ and the independent seven-contrast families at $n=10$ and $n=16$ retain their original intervals; reported estimates and interval endpoints for these contrasts and the secondary scalar gains are supplied in Source Data. The relationships recur under each size's fixed calibration, with magnitudes that depend on the prepared-state and continuation ensembles. Figure~\ref{supp:bankfigure} compares bank sensitivity at all four sizes.

Write $T_m=2\mathbb E[(m-g_m)\delta]$ and $T_r=2\mathbb E[(r-g_r)\delta]$. Except for cancellation-index differences, entries are multiplied by 100 after division by the relevant baseline risk. Negative $r$ contributions in AA are displayed with reversed sign. The phase contrast uses the identity baseline risk as the common denominator. These quantities are neither total MSE reductions nor causal percentages.

The $n=10$ and $n=16$ contrasts confirm positive direct alignment in OR AC and AA, opposing residual alignment and stronger cancellation in AA, attenuation after local-$Z$ rotation, and residual-component information in XOR CC. At $n=10$, local-$Z$ attenuation of $T_m$ accompanies a larger scalar-gain point estimate; at $n=16$, both decrease. Direct alignment and total predictive gain are therefore distinct quantities.

\subsection{Matched diagonal-summary control}
\label{supp:diagonalcontrol}
The Z2 control tests whether compact aggregation of already available diagonal expectations can account for the benefit of M3. With the clause-supported pair set $E$ from Eq.~\eqref{eq:three_summaries}, it is
\begin{equation}
 \mathrm{Z2}=\left(\frac{1}{n}\sum_i\langle Z_i\rangle,\;
 \frac{1}{|E|}\sum_{(i,j)\in E}\langle Z_iZ_j\rangle\right).
 \label{supp:diagonalsummaries}
\end{equation}
The pair average is unweighted: repeated co-occurrence in clauses determines membership in $E$ but gives no additional weight. Both summaries are deterministic functions of $B$, which retains its original mixer-probe and context features.

The retrospective comparisons reused the exact-component cohorts of 256 training and 512 evaluation instances per size at $n=10,12,14,16$, including all four conditions and both learners. We fitted $B+\mathrm{Z2}$ and $B+\mathrm{Z2}+\mathrm{M3}$ models with the original instance-grouped folds, training-only scaling, random streams and hyperparameter grids, and reused the original $B$ and $B+\mathrm{M3}$ predictors unchanged. All new models were fixed before evaluation on the previously examined cohort.

Let $R_{\mathrm{Z2}}$, $R_{\mathrm{M3}}$ and $R_{\mathrm{Z2+M3}}$ denote the risks after adding the indicated summaries to $B$. Table~\ref{supp:diagonalcontroltable} gives the descriptive Z2 gain, $1-R_{\mathrm{Z2}}/R_B$, and two contrasts: the additional M3 gain, $1-R_{\mathrm{Z2+M3}}/R_{\mathrm{Z2}}$, and the M3--Z2 comparison on the common baseline scale, $(R_{\mathrm{Z2}}-R_{\mathrm{M3}})/R_B$. The latter two contrasts for Extra Trees in OR AC and XOR CC retain separate eight-member families for $n=12,14$ and $n=10,16$. All eight 95\% simultaneous intervals in the $n=12,14$ family are positive; six of eight in the $n=10,16$ family are positive. Both intervals containing zero concern XOR CC at $n=10$. Each comparison used 4,000 whole-instance bootstrap replicates, retaining the four depth--generator observations and common resampling weights across conditions and learners within each size; sizes used independent streams. Each study also retains its separate 88-metric secondary family, including Ridge and other-condition comparisons. M3 therefore improves these fitted predictors beyond Z2 in OR AC at all four sizes and XOR CC at $n=12,14,16$; the smaller remaining XOR effect at $n=10$ is less precisely resolved by this comparison.

\begin{table}[htbp]
\centering
\small
\caption{Matched diagonal-summary control for Extra Trees. All entries are multiplied by 100. The Z2 column is a descriptive percentage gain; brackets in the other two columns give 95\% simultaneous intervals within separate eight-contrast families for $n=12,14$ and $n=10,16$. Every risk uses 512 evaluation instances and four repeated observations per instance.}
\label{supp:diagonalcontroltable}
\begin{tabular}{lcrcc}
\toprule
Condition & $n$ & Z2 gain & M3 gain after Z2 & M3--Z2 on $R_B$ scale\\
\midrule
OR AC & 10 & $-0.88$ & 19.64 [13.37, 25.91] & 20.08 [13.37, 26.79]\\
OR AC & 12 & $1.19$ & 28.43 [21.86, 35.01] & 26.93 [20.34, 33.51]\\
OR AC & 14 & $0.45$ & 29.33 [22.85, 35.81] & 29.18 [22.67, 35.69]\\
OR AC & 16 & $2.83$ & 33.79 [27.98, 39.59] & 31.72 [25.98, 37.46]\\
XOR CC & 10 & $1.52$ & 2.35 [-2.96, 7.65] & 4.20 [-1.13, 9.53]\\
XOR CC & 12 & $-0.07$ & 8.22 [3.48, 12.96] & 10.97 [6.38, 15.55]\\
XOR CC & 14 & $0.22$ & 12.57 [9.02, 16.13] & 12.34 [8.39, 16.30]\\
XOR CC & 16 & $4.35$ & 13.88 [7.80, 19.97] & 7.52 [1.59, 13.45]\\
\bottomrule
\end{tabular}
\end{table}

\subsection{Preparation rewards and unperturbed continuation outcomes}
\label{supp:statequality}
We characterized the operating regimes by replaying the states and continuation banks of the exact-component evaluation cohorts at $d=0$. For each state, $Q=\langle R\rangle/(2n)$ is the mean satisfied-clause fraction and $P=\langle\Pi_{\mathrm{good}}\rangle$ is the good-solution probability. The uniform distribution has $Q_{\mathrm{unif}}=7/8$ for OR and $1/2$ for XOR, while $P_{\mathrm{unif}}=\operatorname{Tr}(\Pi_{\mathrm{good}})/2^n$ depends on the instance. Thus mean reward and good-solution probability describe different aspects of the output distribution.

Source Data, \nolinkurl{source_data/preparation_quality_audit.tsv}, reports descriptive means for all four conditions, $n=10,12,14,16$ and both depths. Each row uses 512 independent evaluation instances, shared across conditions at a fixed size; continuation values first average the original 64 draws within an instance. At $d=0$ the intervention generator has no effect, so its two repeated labels were counted only once. The phase control was applied to the prepared state before continuation. No state was selected or excluded according to its reward, and no models were refitted in this audit. The full circuit optimized to select the preparation angles is distinct from these sampled continuations.

For OR, the A-initialized preparation and AA continuation have lower mean rewards and good-solution probabilities than uniform. Switching only the continuation to C raises the mean good-solution probability above uniform at both depths and all four sizes; at $p=8$, this coexists with mean reward below uniform. The local-$Z$ transform preserves prefix reward and good-solution probability but changes continuation outcomes. XOR CC has mean prefix and continuation rewards above uniform at every size. These diagnostics distinguish preparation quality from response predictability, using uniform sampling as a reference distribution.

\subsection{High-reward preparation controls}
\label{supp:highreward}

Table~\ref{supp:highrewardtable} complements Fig.~\ref{fig:highreward} with paired continuation contrasts under $B_{\mathrm{common}}$ (Methods~\ref{meth:highreward}). Full-precision preparation rewards, good-solution probabilities, gains, losses and decomposition terms are supplied in \texttt{source\_data/high\_reward\_preparation\_all\_sizes.tsv}.

\begin{table}[htbp]
\centering
\small
\setlength{\tabcolsep}{4pt}
\caption{Observable gains in high-reward preparations and paired continuation contrasts. Entries are estimates [95\% simultaneous intervals]. At each size, the upper two primary comparisons and lower two secondary comparisons form separate families. Gains are Extra Trees MSE reductions relative to the common baseline; pp denotes percentage points; $C$ is the physical cancellation index in Eq.~\eqref{eq:cancellation}.}
\label{supp:highrewardtable}
\begin{tabular}{lcccc}
\toprule
Comparison & $n=10$ & $n=12$ & $n=14$ & $n=16$\\
\midrule
\multicolumn{5}{l}{\textit{Primary: predictive gains with H preparation}}\\
H+A gain (\%) & $-1.51\;[-3.77,\,0.74]$ & $1.58\;[-0.15,\,3.32]$ & $-0.47\;[-2.37,\,1.43]$ & $0.45\;[-1.41,\,2.32]$\\
H+C gain (\%) & $1.20\;[-0.82,\,3.21]$ & $6.89\;[4.47,\,9.32]$ & $5.81\;[3.10,\,8.53]$ & $8.39\;[5.50,\,11.28]$\\
\midrule
\multicolumn{5}{l}{\textit{Secondary: continuation differences at fixed H preparation}}\\
Gain C$-$A (pp) & $2.71\;[-0.41,\,5.83]$ & $5.31\;[2.32,\,8.30]$ & $6.28\;[2.96,\,9.60]$ & $7.94\;[4.53,\,11.34]$\\
$C_{\mathrm{HA}}-C_{\mathrm{HC}}$ & $0.347\;[0.257,\,0.437]$ & $0.867\;[0.806,\,0.927]$ & $0.955\;[0.912,\,0.998]$ & $0.809\;[0.774,\,0.845]$\\
\bottomrule
\end{tabular}
\end{table}

The loss decomposition further resolves the retained information: direct and complementary residual alignments are both positive in the H+C Extra Trees accounts at $n=12,14,16$. At $n=14,16$, the weak H+A gain accompanies smaller alignment terms, contrasting with the large opposing terms of the exact-component AA regime.

The H+C baseline MSE is about $3.3$ times its A+C value at $n=12$ and $14$, and $3.8$ times at $n=16$. These paired descriptive risk changes explain why the relative gains and the retained absolute reductions in Fig.~\ref{fig:highreward}c convey different aspects of the preparation comparison.

\subsection{Robustness across evaluation and training samples at \texorpdfstring{$n=10$}{n=10}}
\label{supp:n10robustness}
On 1,024 fresh evaluation instances, both primary XOR comparisons and all five independently trained cohorts have positive simultaneous bounds in their respective families (Table~\ref{tab:n10-robustness}; Methods~\ref{meth:n10}). The five-cohort estimator averages losses before taking their ratio; a descriptive crossed bootstrap of training cohorts and evaluation instances gives a 3.86--8.62\% interval. The frozen exact-component model's gain exceeds its estimate on the exact-component evaluation cohort by 2.57 percentage points, with a descriptive pointwise interval of $[-2.24,7.55]$ percentage points. Source Data includes the corresponding results for all four circuit conditions.

\begin{table}[htbp]
\centering
\small
\setlength{\tabcolsep}{10pt}
\caption{XOR sample robustness at $n=10$ on 1,024 fresh evaluation instances. Entries are Extra Trees/M3 MSE reductions (\%) and 95\% simultaneous intervals. The upper two comparisons retain the original two-contrast primary family. The five individual-cohort comparisons retain their original 54-contrast secondary family, which also contains the other circuit conditions reported in Source Data. Each new training cohort contains 256 instances.}
\label{tab:n10-robustness}
\begin{tabular}{lrrr}
\toprule
XOR CC predictor & Gain (\%) & Lower & Upper\\
\midrule
\multicolumn{4}{l}{\textit{Primary comparisons}}\\
Frozen exact-component model & 8.29 & 5.29 & 11.29\\
Five-cohort mean risk & 6.18 & 4.29 & 8.08\\
\midrule
\multicolumn{4}{l}{\textit{Individual new training cohorts (secondary family)}}\\
Cohort 1 & 4.85 & 0.38 & 9.31\\
Cohort 2 & 3.52 & 0.90 & 6.14\\
Cohort 3 & 8.13 & 3.60 & 12.65\\
Cohort 4 & 7.16 & 2.89 & 11.44\\
Cohort 5 & 7.45 & 4.91 & 9.98\\
\bottomrule
\end{tabular}
\end{table}

\subsection{Robustness under matched calibration budgets}
\label{supp:calibrationrobustness}
The matched-budget study compares the original C with three independently calibrated C sets at $n=14,16$ on paired fresh cohorts (Methods~\ref{meth:calibrationcentres}). Table~\ref{supp:calibrationrobustnesstable} reports the Extra Trees/M3 gains within their separate inference families.

\begin{table}[htbp]
\centering
\small
\setlength{\tabcolsep}{8pt}
\caption{Extra Trees/M3 prediction-MSE reductions (\%) in the calibration-budget study. All rows use the same fresh 512 evaluation instances per size. C$_1$--C$_3$ each use an independent 64-instance calibration cohort, 100 updates per restart and two restarts. Intervals are 95\% simultaneous bounds in separate families: six new-C OR gains, six new-C XOR gains and four original-C comparators. Original-C rows are fresh-cohort estimates, distinct from Fig.~\ref{fig:gain}.}
\label{supp:calibrationrobustnesstable}
\begin{tabular}{clrrrr}
\toprule
$n$ & Calibration & OR AC & 95\% interval & XOR CC & 95\% interval\\
\midrule
14 & Original C & 27.70 & [21.61, 33.79] & 11.56 & [7.50, 15.62]\\
14 & C$_1$ & 29.53 & [23.70, 35.35] & 12.96 & [8.34, 17.58]\\
14 & C$_2$ & 29.87 & [24.01, 35.73] & 11.22 & [6.94, 15.51]\\
14 & C$_3$ & 26.11 & [20.04, 32.18] & 12.22 & [8.47, 15.97]\\
\midrule
16 & Original C & 33.35 & [26.78, 39.91] & 10.04 & [5.27, 14.81]\\
16 & C$_1$ & 33.22 & [26.77, 39.68] & 4.44 & [-0.80, 9.68]\\
16 & C$_2$ & 33.12 & [26.62, 39.61] & 11.35 & [6.80, 15.91]\\
16 & C$_3$ & 33.49 & [27.11, 39.87] & 7.09 & [1.88, 12.31]\\
\bottomrule
\end{tabular}
\end{table}

All six new-C OR gains have positive simultaneous bounds. Their paired differences from the original C range from $-1.60$ to $+2.17$ percentage points at $n=14$ and $-0.23$ to $+0.14$ at $n=16$; all six difference intervals include zero. These results support persistence of the OR advantage under the expanded budget, without requiring equality between calibrations. For XOR, five of six new-C gains have positive simultaneous bounds. At $n=16$, C$_1$ gives 4.44\% [$-0.80,9.68$]\%, and its paired change from the original C is $-5.60$ percentage points [$-9.68,-1.51$]. Its absolute MSE reduction also decreases. The XOR results thus resolve calibration-dependent variation alongside retained benefits at the other new centres. All sixteen Ridge gains have positive simultaneous bounds in their separate sensitivity family.

Full-precision risks, preparation diagnostics and all contrasts are supplied in \nolinkurl{source_data/calibration_robustness_results.json} and \nolinkurl{source_data/calibration_robustness_contrasts.tsv}; \nolinkurl{source_data/calibration_robustness_schedules.json} contains the new angles. This scalar-response study complements the original exact-component cohorts, whose physical decompositions retain the original schedules.

\end{document}